\documentclass[a4paper,11pt]{article}
\pdfoutput=1 

\usepackage{jheppub} 

\usepackage[T1]{fontenc} 

\subheader{CERN-PH-TH-2015-014, DFPD-2015/TH/05}

\title{\boldmath Robust collider limits on \\ heavy-mediator Dark Matter}

\author[a]{Davide Racco,}
\author[b]{Andrea Wulzer,}
\author[\, b,c]{and Fabio Zwirner}

\affiliation[a]{
D\'epartement de Physique Th\'eorique and Center for Astroparticle Physics, \\ 
Universit\'e de Gen\`eve, 24 quai Ansermet, CH-1211 Gen\`eve 4, Switzerland}
\affiliation[b]{Dipartimento di Fisica e Astronomia `G.~Galilei', Universit\`a di Padova 
\\ 
and INFN, Sezione di Padova, Via Marzolo 8, I-35131 Padova, Italy}
\affiliation[c]{Physics Department, Theory Unit, CERN, CH-1211 Gen\`eve 23, Switzerland}

\emailAdd{davide.racco@unige.ch}
\emailAdd{andrea.wulzer@pd.infn.it}
\emailAdd{fabio.zwirner@pd.infn.it}

\newcommand{\Mmed}{M_\text{med}}
\newcommand{\Gmed}{\Gamma_\text{med}}
\newcommand{\MDM}{m_\text{DM}}
\newcommand{\Mcut}{M_\text{cut}}
\newcommand{\MET}{E_\text{T}^\text{miss}}
\newcommand{\pTjet}{p_\text{T}^\text{jet}}
\newcommand{\pjet}{p^\text{jet}}
\newcommand{\Ecm}{E_\text{cm}}%
\newcommand{\Qtr}{Q_\textrm{tr}}
\newcommand{\Mblind}{M_*^\text{EFT}}
\newcommand{\Mresc}{\widetilde M_*}
\newcommand{\mP}{M_\text{P}} 
\newcommand{\ODM}{\Omega_\text{DM}}
\newcommand{\mDM}{m_\text{DM}}
\newcommand{\mZp}{m_{Z^{\, \prime}}}
\newcommand{\GZp}{\Gamma_{Z'}}

\newcommand{\msq}{\widetilde m}
\newcommand{\Gsq}{\Gamma_{\widetilde q}}
\newcommand{\gDM}{g_\text{DM}}

\abstract{We discuss how to  consistently use Effective Field Theories (EFTs) to set
universal bounds on heavy-mediator Dark Matter at colliders, without prejudice on the model underlying a given effective interaction.  
We illustrate the method for a Majorana fermion, universally coupled to the Standard Model quarks via a dimension-6 axial-axial four-fermion operator.
We recast the ATLAS mono-jet analysis and show that a considerable fraction of the parameter space, seemingly excluded by a na\"ive EFT interpretation, is actually still unexplored. 
Consistently set EFT limits can be reinterpreted in any specific underlying model. 
We provide two explicit examples for the chosen operator and compare the reach of our model-independent method with that obtainable by dedicated analyses.
}

\begin{document} 
\maketitle
\flushbottom

\section{Introduction}
\label{sec:intro}

After the discovery \cite{hdisc-atlas,hdisc-cms} of a Higgs boson compatible \cite{hprop-atlas,hprop-cms}  with the Standard Model (SM), and the non-detection so far of new particles \cite{bsmlhc} at the LHC, searches for Dark Matter (DM) in the form of a Weakly Interacting Massive Particle (WIMP) are becoming a central theme for the LHC general purpose experiments (for a pedagogical review, see e.g.~\cite{dmlhcslac}).

In the WIMP hypothesis, the DM particle has a mass in the GeV--TeV range, and the strength of its couplings to the SM particles is roughly of electroweak size. 
The relic density, generated by thermal freeze-out, can then match the cosmological and astrophysical observations. 
This so-called ``WIMP miracle'' receives further support from the fact that WIMPs are ubiquitous in new physics models of ElectroWeak Symmetry Breaking (EWSB), motivated by the naturalness problem of the SM. 
The latter is an appealing and intensively explored possibility, but WIMP DM might well originate from a completely unrelated sector. 
Moreover, we currently have no idea of how the complete EWSB sector looks like, thus there is not much we can say a priori on the specific properties of WIMP DM. 

In the situation described above, a general and model-independent exploration appears mandatory.
Commitment to specific benchmark models (or classes of benchmark models) should be avoided whenever possible in the analysis of experimental data, or at least treated as an accessory step in the interpretation. 
The goal is to search for WIMP DM in a comprehensive way, leaving no unexplored corners in theory space.\footnote{For example, when planning future direct-detection experiments sensitive to low-mass WIMPs in the 1--10 GeV range, it may be important to understand on general grounds how much room for discovery is left after the so far unsuccessful LHC searches, without being committed to specific benchmark models.}

In the case of heavy-mediator DM, this program can be carried out, at least to some extent. 
The working hypothesis is that the DM candidate $X$ interacts with the SM through the exchange of one or more particles, called ``mediators'', whose mass is well above the mass $\MDM$ of the DM particle. 
This assumption is motivated by the present lack of evidence for new particles at the LHC, but it is not the only possibility. 
The case in which the mediator is a SM particle, such as a weak or the Higgs boson, is equally plausible and deserves equal attention. 
Light and very weakly coupled mediators can be also conceived.

Focusing on the heavy-mediator case for the rest of this paper, it is relatively easy to set up a model-independent strategy for DM searches. We can exploit the fact that the dynamics of the DM particle can be universally described, in the appropriate kinematical regime, by a low-energy EFT Lagrangian \cite{DMEFT, 0403004, 0808.3384, 0912.4511, 1002.4137, 1005.1286, 1005.3797, 1008.1783, 1108.1196, 1109.4398, 1402.1173, 1403.5161, 1411.3342, 1412.0520}, invariant under the SM gauge group and the Lorentz group:\footnote{At energies as low as those relevant for direct detection experiments, it may even be convenient to switch to a non-relativistic EFT \cite{1008.1591, 1203.3542, 1211.2818, 1307.5955}, but for obvious reasons this approach precludes a direct comparison with collider searches and will not be pursued here.}
\begin{equation}
\label{LEFT}
{\cal L}_{\textrm{EFT}} = {\cal L}_{\textrm{SM}} + {\cal L}_X + {\cal L}_{\textrm{int}} \, .
\end{equation}
In the above equation, ${\cal L}_{\textrm{SM}}$ denotes the SM Lagrangian, ${\cal L}_X$ is the free Lagrangian for $X$, and ${\cal L}_{\textrm{int}}$ contains the operators describing the DM interactions with the SM particles, plus possible additional interactions in the DM and SM sectors. 
If we knew the true microscopic DM theory, these operators could be computed by integrating out the mediators.
However, their form is universal and the lack of information on the mediator dynamics merely prevents us from computing the value of their coefficients, which are thus free input parameters of the EFT. 

The allowed operators in ${\cal L}_{\textrm{int}}$ can be classified according to their mass dimension $d$, for different hypotheses on the DM quantum numbers.
In many relevant cases the DM quantum numbers forbid renormalizable interactions with $d \le 4$, and the lowest-dimensional operators have $d=5,6$.
For the physics to be considered in this paper, we can assume that the $d=5$ operators are negligible and the leading operators have $d=6$:   
\begin{equation}
\label{LINT}
 {\cal L}_{\textrm{int}}=\frac1{M_*^2}\sum_i c_i \, O_i \, ,
\end{equation}
where the sum runs over all $d=6$ operators $O_i$ allowed by the symmetries, $c_i$ are dimensionless coefficients and the overall effective coupling strength is parameterized by a dimensionful coupling $1/M_*^2$. 

While the EFT can be formally defined independently of any consideration about its microscopic origin, its range of applicability and thus its physical relevance depend on the underlying theory.
Namely, the EFT provides an accurate description of the underlying model only for elementary scattering processes taking place at a low enough centre-of-mass energy $E_{\textrm{cm}}$, below a certain critical scale $\Mcut$ usually called the EFT cutoff. 
This cutoff is determined by the mass of the mediators in the microscopic theory but it is unknown from the viewpoint of the EFT and it should thus be treated as a free parameter, on the same footing as those introduced above. 
\newpage 

The EFT is then characterised by at least three parameters:~
\begin{itemize}
\vspace{-7pt}
\item the DM mass $\MDM$;
\vspace{-7pt}
\item the scale $M_*$ of the interaction;
\vspace{-7pt}
\item the cutoff scale $\Mcut$.
\vspace{-7pt}
\end{itemize}
If a single operator appears in eq.~(\ref{LINT}), the corresponding dimensionless coefficient  can be absorbed in $M_*$, otherwise the EFT parameters also include the $c_i$ coefficients. 
With these free parameters, the EFT faithfully reproduces the predictions of any microscopic theory for all processes taking place at $E_{\textrm{cm}}<\Mcut$. 
Given that the effective operators in eq.~(\ref{LINT}) may have many possible microscopic origins, exemplified by the plethora of models in the literature, this simplification is particularly useful. 

Notice that $\Mcut$ and $M_*$ are logically independent parameters, however they can be approximately related by
\begin{equation}
\label{wittenpolchinski}
\Mcut=g_* \, M_*\,,
\end{equation}
where $g_*$ is a suitably defined coupling strength of the underlying microscopic theory. 
The simplest way to motivate the above equation is the analogy with the Fermi theory of weak interactions, where the cutoff $\Mcut$ is the mass of the $W$ boson (the mediator in this context), $g_*$ is the $SU(2)$ gauge coupling $g_w$ and $1/M_*^2$ is the Fermi constant $G_F$: they indeed obey eq.~(\ref{wittenpolchinski}) up to numerical factors. 
Alternatively, the physical meaning of $g_*$ can be appreciated by noticing that the EFT interaction strength is given, for processes taking place at a given center-of-mass energy $E_{\textrm{cm}}$, by the dimensionless combination $E_{\textrm{cm}}^2/M_*^2$. 
At the mediator scale, i.e. the cutoff scale $\Mcut$, this strength becomes $\Mcut^2/M_*^2=g_*^2$, providing further justification for interpreting $g_*$ as the typical mediator coupling. 
Using eq.~(\ref{wittenpolchinski}) to re-express $M_*$ in terms of $g_*$ will be important in section~2.2, in order to draw the current limits on a plane suited for theoretical interpretation.

The EFT can be straightforwardly used to predict the cross-sections for a number of relevant reactions, namely the DM annihilation in the Early Universe, which determines the thermal relic density, the present-day annihilation, which controls indirect detection, and the DM scattering on nucleons, which direct search experiments try to detect. 
Indeed, all these reactions take place at safely small $E_{\textrm{cm}}$ and therefore, up to subtle effects that might be encountered in the relic density calculation, the EFT predictions are automatically trustable. 
If collider searches could be added to the list, we would reach the truly remarkable conclusion that all the experimental information on heavy-mediator DM can be simultaneously interpreted and compared in a completely model-independent fashion, with no prejudice  on the specific nature of the mediator and of its couplings to DM and to the SM. 
However, the usage of the EFT at colliders is problematic, because the energy of the reaction in which the DM is produced is not necessarily smaller than $\Mcut$, and this risks to invalidate the EFT predictions. 
The effect is quantitatively amplified by the requirement of extra hard objects (e.g., one jet), in addition to the pairs of DM particles, for the signal to be triggered and disentangled from the background. 
This problem has been discussed at length in the recent literature (see e.g. refs.~\cite{Fox, 1111.5331, riottos1, BDM, riottos2, riottot, Englert:2014cva, AABB, Malik, Buckley:2014fba}), the goal of the present article is to illustrate a simple and practical solution.

The basic observation is that the processes for DM production at colliders can be split into two kinematically distinct classes, characterised by a centre-of-mass energy below and above $\Mcut$, respectively. 
The former class defines our theoretical signal, and its rate is accurately predicted by the EFT. 
The latter would instead require the knowledge of the microscopic theory and its contribution to the cross-section is thus unpredictable within the EFT.
Under certain conditions, to be described below, the second class can be simply ignored and an experimental limit can be set on the signal defined, as explained above, by the DM production reaction restricted to \mbox{$E_{\textrm{cm}} < \Mcut$}. 
This is possible if the experimental search is performed as a counting experiment in one or several signal regions, defined by a certain set of cuts on the visible final state particles. 
The low and high $E_{\textrm{cm}}$ processes both contribute to each signal region, but in a purely additive way, since low and high $E_{\textrm{cm}}$ regions are quantum-mechanically distinguishable and do not interfere. Therefore a lower bound on the expected cross-section is obtained by considering only the ``well-predicted'' signal events, namely those restricted to the \mbox{$E_{\textrm{cm}}<\Mcut$} region. 
If the result of the search is negative, an exclusion upper bound $\sigma_{\textrm{exc}}$ is set on the cross-section, which we can interpret through the inequality
\begin{equation}
\label{sigdef}
{\sigma}^S_{EFT} \Bigr\rvert_{E_{\textrm{cm}}<\Mcut} \le {\sigma}_{\textrm{true}}^S<\sigma_{\textrm{exc}} \, ,
\end{equation}
where ${\sigma}_{\textrm{true}}^S$ denotes the ``true'' signal as it would be computed in the unknown microscopic theory. 
Regardless of what the latter theory is, the restricted EFT signal ${\sigma}^S_{EFT}$ systematically underestimates the cross-section and thus provides a conservative, but correct, exclusion limit.\footnote{For a similar approach in the context of Higgs EFTs, see \cite{Riva:2014}.}

The rest of the paper is organised as follows. In section~2 we illustrate our limit-setting strategy in the explicit example of a four-fermion operator obtained as the product of axial currents involving the SM quarks and a SM-singlet Majorana fermion DM. 
This choice is partly motivated by the fact that direct and indirect detection experiments have a poor sensitivity to this operator, thus collider searches are expected to be the most sensitive ones, but the same method can be applied to all other operators.
We quantify the reach of current collider searches by recasting the ATLAS mono-jet results available at the time of this work, and show how the latter can be presented in a theoretically useful way. 
Besides the methodological proposal, the important physics point is that, from the general EFT viewpoint, the present collider bounds on DM have not yet probed the most plausible region of parameter space. 
To access such region, we need not only more energy and luminosity, as expected in the forthcoming runs of the LHC, but also improvements in the experimental analyses. 
In section~3 we describe another relevant feature of our strategy, the fact that the limits set in the EFT can be straightforwardly re-interpreted as constraints on any specific microscopic model. 
This is because the EFT parameters can be computed in the underlying microscopic theory and expressed in terms of the ``fundamental'' parameters of the latter (for previous discussions of the interplay between EFT and underlying microscopic models in DM searches at colliders, see again refs.~\cite{Fox, 1111.5331, riottos1, BDM, riottos2, riottot, Englert:2014cva, AABB, Malik, Buckley:2014fba}). 
We consider two representative models, Model A and Model B, which both give rise to the same axial-axial effective operator, and compare the limits derived from the EFT with those obtainable from a dedicated interpretation of the mono-jet search within the two models. 
Since our signal cross-section systematically underestimates that of the microscopic theory, we obtain conservative limits.
We find that these limits differ significantly from those obtained in the full models only in the kinematical region where the mediators can be resonantly produced. 
In such a case, however, more comprehensive experimental strategies, complementing the event selection used for heavy-mediator DM searches with other selections that can take full advantage of the resonant production of the mediators (single or in pairs, with one or more jets in the event), should be able to provide stronger bounds. 
We end this section by discussing two aspects of our simple and practical approach that can be helpful for the comparison with a similar but more model-dependent approach recently put forward in \cite{riottos1, riottos2, riottot}. 
We finally present our conclusions in Section~4.
Some back-up material is collected in three appendices. 
Appendices~A and B provide details on Model~A and Model~B, respectively.
Appendix C collects the approximate analytical formulae used to draw the relic density constraint in some of the figures. 

\section{Limit-setting strategy}
\label{sec:limsetstr}

For the present study, we assume that the DM particle is a Majorana fermion, singlet under the SM gauge group and represented by a self-conjugate four-component spinor $X = X^c$, whose free Lagrangian reads
\begin{equation}
\label{freeX}
{\cal L}_X = \frac 12 \, \overline{X} \, (i \partial \!\!\! \slash - \MDM) \, X \, .
\end{equation}

As for the interactions between $X$ and the SM particles, we just choose a representative example to illustrate our limit-setting strategy, assuming that they can be described, in the low-energy limit, by the single~\footnote{Radiative corrections may generate additional operators \cite{haisch,deramo}, this can be important when comparing with direct dark matter searches but does not play a role in the present context.} axial-axial four-fermion operator~\footnote{This operator is twice the M6 operator in \cite{1005.1286}, and formally coincides with the D8 operator in \cite{1008.1783}, which is often taken as a benchmark for experimental searches. Notice however that we are dealing with a Majorana spinor normalised as in (\ref{freeX}), while D8 involves a canonically normalised Dirac spinor.}
\begin{equation}
O = - \frac{1}{M_*^2} \, \left( \overline X \gamma^\mu \gamma^5 X \right) \left( \sum_q \overline q\gamma_\mu \gamma^5 q \right) \, ,
\label{eq:EFTlag}
\end{equation}
where the sum is over all quark flavours ($q=u,d,c,s,t,b$), the dimensionless coefficient $c$ has been re-absorbed in the definition of $M_*$, and the overall minus sign is purely conventional in the present context. 
This effective operator mediates DM pair-production at the LHC, a process which is however undetectable and impossible to trigger because of the lack of visible objects in the final state. 
Searches are performed by considering extra visible emissions from the initial quarks, leading to the so-called ``mono-$V$'' signatures, where $V$ could be a jet  \cite{Chatrchyan:2011nd, Aad:2011xw, Aad:7tevmonojet, Aad:8tevmonojet, Chatrchyan:2012me, Khachatryan:2014rra}, a photon \cite{Chatrchyan:2012tea, Aad:2012fw, Aad:2014tda, Khachatryan:2014rwa}, a massive weak boson \cite{Aad:2013oja, Aad:2014vka} or a top quark \cite{Aad:2014wza, Aad:2014vea}. 
Below we restrict our attention to the mono-jet searches, because they currently show the best sensitivity, but our considerations also apply to the other channels.

\subsection{ATLAS mono-jet recast}
\label{ATMR}

Searches for a jet plus missing transverse energy ($\MET$) have been performed at the LHC by the ATLAS \cite{Aad:2011xw,Aad:8tevmonojet,Aad:7tevmonojet} and CMS \cite{Chatrchyan:2011nd,Chatrchyan:2012me,Khachatryan:2014rra} collaborations.\footnote{For a very recent update of the ATLAS mono-jet analysis, which appeared after the completion of our work, see also \cite{Aad:2015zva}.} 
We focus here on the ATLAS analysis in ref.~\cite{Aad:8tevmonojet} because, among those available at the time of the present work, it is particularly suited to illustrate the general point we would like to make.
The search is performed as a counting experiment in four overlapping signal regions (SR), with pre-selected events characterized by \mbox{$\MET>120$~GeV}, one jet with \mbox{$\pTjet>120$~GeV}, \mbox{$|\eta|<2$} and at most one additional jet with \mbox{$p_\text{T}>30$~GeV} and \mbox{$|\eta|<4.5$}. 
If found, the second jet is asked to be separated in the azimuthal direction from the $\vec p_\text{T}^\text{\ miss}$ by a cut $\Delta\phi > 0.5$. 
Additional requirements, namely on the primary vertex reconstruction and on the absence of extra jets with anomalous charged/calorimetric composition, are not directly relevant for our study, since their impact crucially depends on the detector response, which we cannot simulate. 
The four signal regions \mbox{SR$i$} ($i=1,2,3,4$) are defined by increasingly strong cuts on $\MET$ and on $\pTjet$. 
The results are presented as upper bounds, $\sigma^i_{\textrm{exc}}$, on the visible cross-section in each region. 
The SR definitions and the exclusion limits are summarized in table~\ref{tab:resATLAS}. 
\begin{table}
\centering
\begin{tabular}{ l   c  c  c  c  }
\hline
signal region & SR$1$ & SR$2$ & SR$3$ & SR$4$  \\
$\pTjet$ and $\MET$ & \mbox{$>\hspace{-3pt}120$} & $>\hspace{-3pt}220$ & $>\hspace{-3pt}350$ & $>\hspace{-3pt}500$  \\
$\sigma_{\textrm{exc}}$[pb] & \;\;$2.7$\;\; & \;$0.15$\; & $4.8\, 10^{-2}$ & $1.5\, 10^{-2}$ \\
\hline
\end{tabular}
\caption{Signal region definitions (cuts expressed in GeV) and $95\%$~CL limits from ref.~\cite{Aad:8tevmonojet}. }
\label{tab:resATLAS}
\end{table}

We reinterpret these limits as follows. The expected signal in each SR is expressed as 
\begin{equation}
\sigma_{\textrm{SR}i}=\sigma\times A_i \times\epsilon_i\,,
\end{equation}
where $\sigma$ denotes the total signal cross-section defined as in eq.~(\ref{sigdef}), $A_i$ is the geometric cut acceptance, as obtained from a leading-order parton-level simulation, and the efficiency $\epsilon_i$ is the correction due to showering, hadronization and detector effects. 
Acceptances and efficiencies depend on the DM mass $\MDM$ and on the cutoff $M_{\textrm{cut}}$, while the operator scale $M_*$ only enters the total cross-section as an overall factor $1/M_*^4$. 
We compute the parton-level quantities $\sigma$ and $A_i$ by \texttt{MadGraph~5} \cite{madgraph5} simulations, while we estimate the $\epsilon_i$ corrections by matching with the limits on the \mbox{D$8$} operator scale reported in ref.~\cite{Aad:8tevmonojet}. 
In practice, we simulate the same \mbox{D$8$} operator signal considered in ref.~\cite{Aad:8tevmonojet} (i.e. $M_{\textrm{cut}}=\infty$ in eq.~\eqref{sigdef}), we compute $\sigma\times A_i$ and we determine $\epsilon_i$ such as to reproduce the ATLAS limit on the effective operator scale as a function of the DM mass.
Actually, since only the third SR is used by ATLAS to set the limit, only $\epsilon_3$ can be obtained in this way. 
The same efficiencies are used for the  other SRs, although we see no reason why the efficiency should stay the same in all the regions. 
The result of this procedure gives rather small efficiencies, of around $60\%$, approximately constant over the whole DM mass range. 
We verified that this considerable signal loss is mainly due to the fact that our simulation does not include the showering-level production of extra jets, a significant fraction of which are vetoed in the ATLAS event selection. 
Notice that the efficiencies for our signal might be significantly different from those estimated in the na\"ive EFT because, although based on the same effective operator \mbox{D$8$} of eq.~(\ref{eq:EFTlag}), our signal is constrained by $M_{\textrm{cut}}$ to the low invariant mass region, thus it is expected to have different kinematical distributions. 
A complete simulation in different regions of $\MDM$ and $M_{\textrm{cut}}$, including showering and matching, would be needed for an accurate analysis, but goes beyond the aim of the present illustrative example.

Under the assumptions explained above, the expected signal takes the form
\begin{eqnarray}
\label{xsectpar}
\displaystyle
\sigma_{\textrm{SR}i} (M_*,\MDM,M_{\textrm{cut}})
=&&
\sigma (M_*,\MDM,M_{\textrm{cut}}) \times A_i(\MDM,M_{\textrm{cut}}) \times \epsilon \nonumber \\
=&&
\left[\frac{1\,\textrm{TeV}}{M_*}\right]^4 \times \overline\sigma (\MDM,M_{\textrm{cut}}) \times A_i(\MDM,M_{\textrm{cut}}) \times \epsilon \,,
\end{eqnarray}
where the overall scaling of the cross section with $M_*$ has been factored out and the result expressed in terms of a reference cross-section $\overline\sigma$ computed for $M_*=1$~TeV. 
The reference cross-section times the acceptances are obtained by \texttt{MadGraph~5} \cite{madgraph5} simulations of DM pair plus one parton production, duly restricted by the hard jet kinematical cuts that define each SR. 
$\MET$ cuts are automatically imposed because the jet and the missing transverse momentum, i.e. the transverse momentum of the DM pair, are back-to-back in our parton-level sample. 
The theoretical restriction $E_{\textrm{cm}}<M_{\textrm{cut}}$, which ensures the validity of the EFT description as explained in the introduction, should be imposed as a cut on the total invariant mass of the hard final states of the reaction, namely as
\begin{equation}
\label{eq:Mcut}
\displaystyle
\left[p(DM_1)+p(DM_2)+\pjet\right]^2<M_{\textrm{cut}}^2\,.
\end{equation}
For our parton level simulation this is equivalent to a cut $\sqrt{\widehat{s}}<M_{\textrm{cut}}$ on the total partonic centre-of-mass energy, however when going to the showered and matched level one should be careful not to cut on $\sqrt{\widehat{s}}$ but on the variable in eq.~(\ref{eq:Mcut}), with $\pjet$ the leading jet four-momentum.

A scan is performed in the $(\MDM,M_{\textrm{cut}})$ plane for each SR and the values of $\sigma\times A_i$ are used to construct two-dimensional interpolating functions.
A significant dependence on $\MDM$ is only found for $\MDM\gtrsim 80$~GeV, while for smaller values $\sigma\times A_i$ is basically constant in $\MDM$. 
Once the signal cross-sections are known, the $95\%$~CL limits are imposed as constraints
\begin{equation}
\label{limits}
\sigma_{\textrm{SR}i} (M_*,\MDM,M_{\textrm{cut}})<\sigma_{\textrm{exc}}^i\,,
\end{equation}
out of which the $95\%$~CL allowed  regions are immediately found in the three-dimensional parameter space $(M_*,\MDM,M_{\textrm{cut}})$. 
The limits from the various signal regions can be studied separately or combined.
For our illustrative purposes, the combination will be performed by just taking the overlap of the four allowed regions.  
The results of this simple limit-setting procedure are discussed in the following section.

\subsection{Results and discussion}

\begin{figure}[t]
  \centering
  \includegraphics[width=0.49\textwidth]{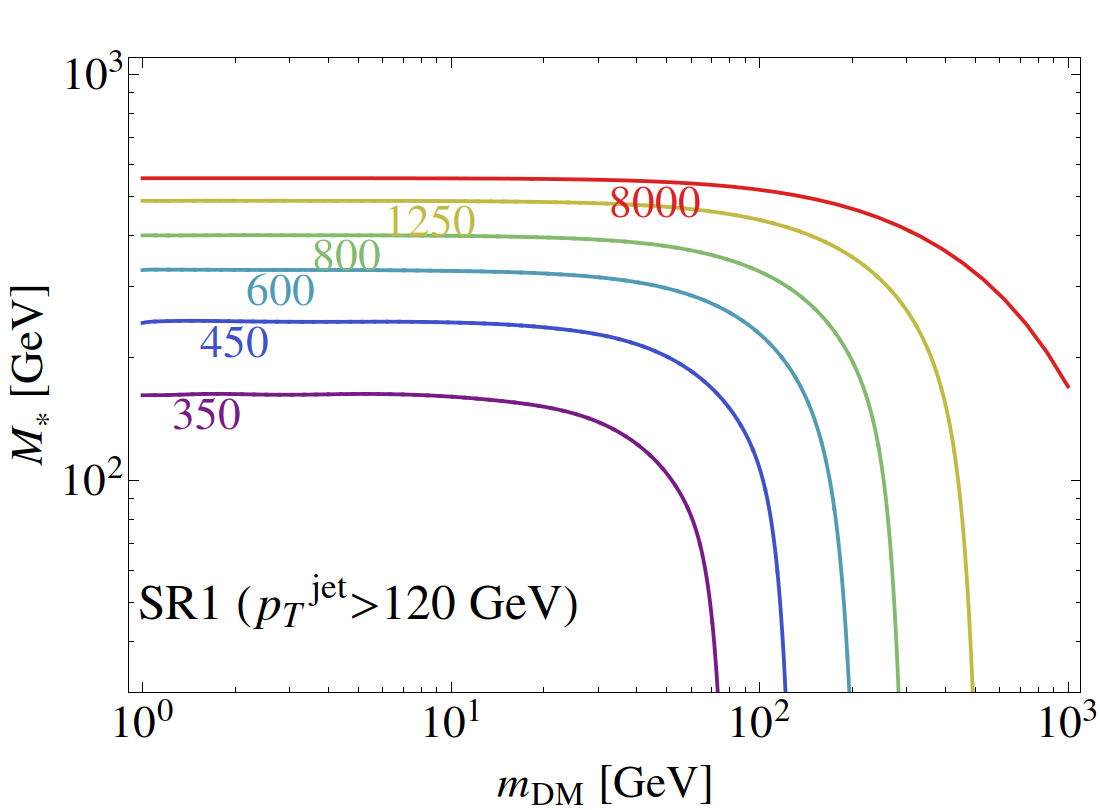} \hfill
  \includegraphics[width=0.49\textwidth]{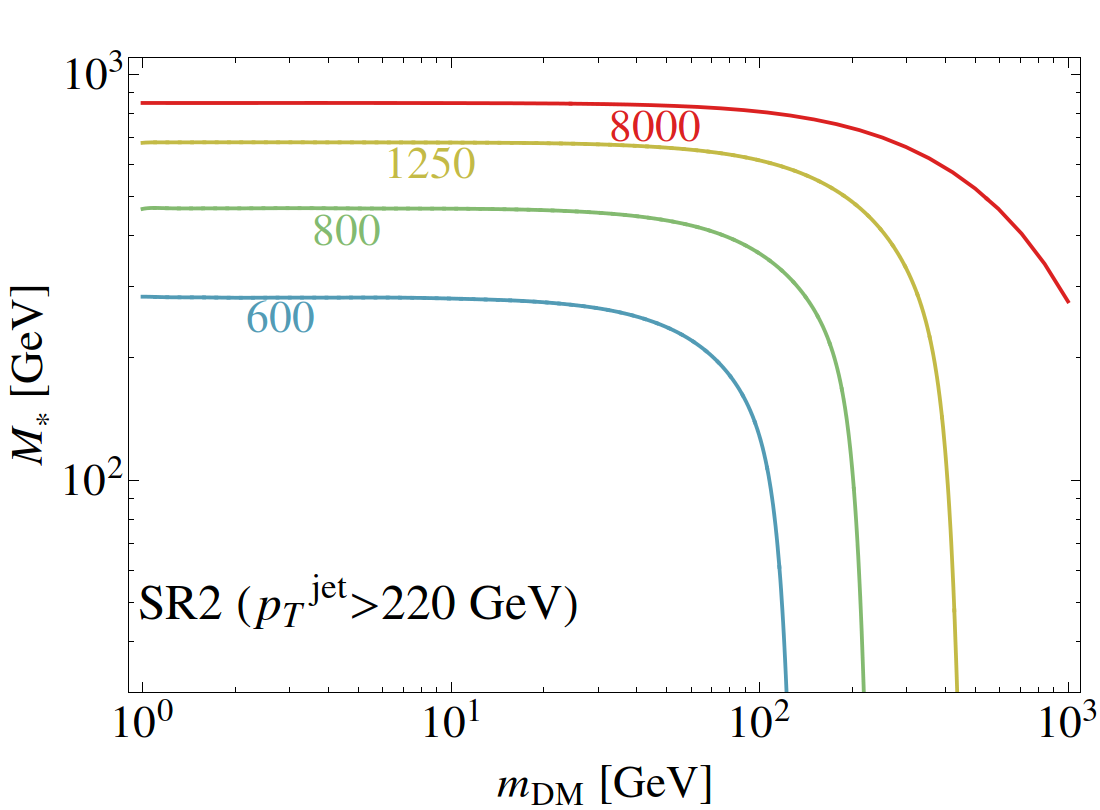}
  \includegraphics[width=0.49\textwidth]{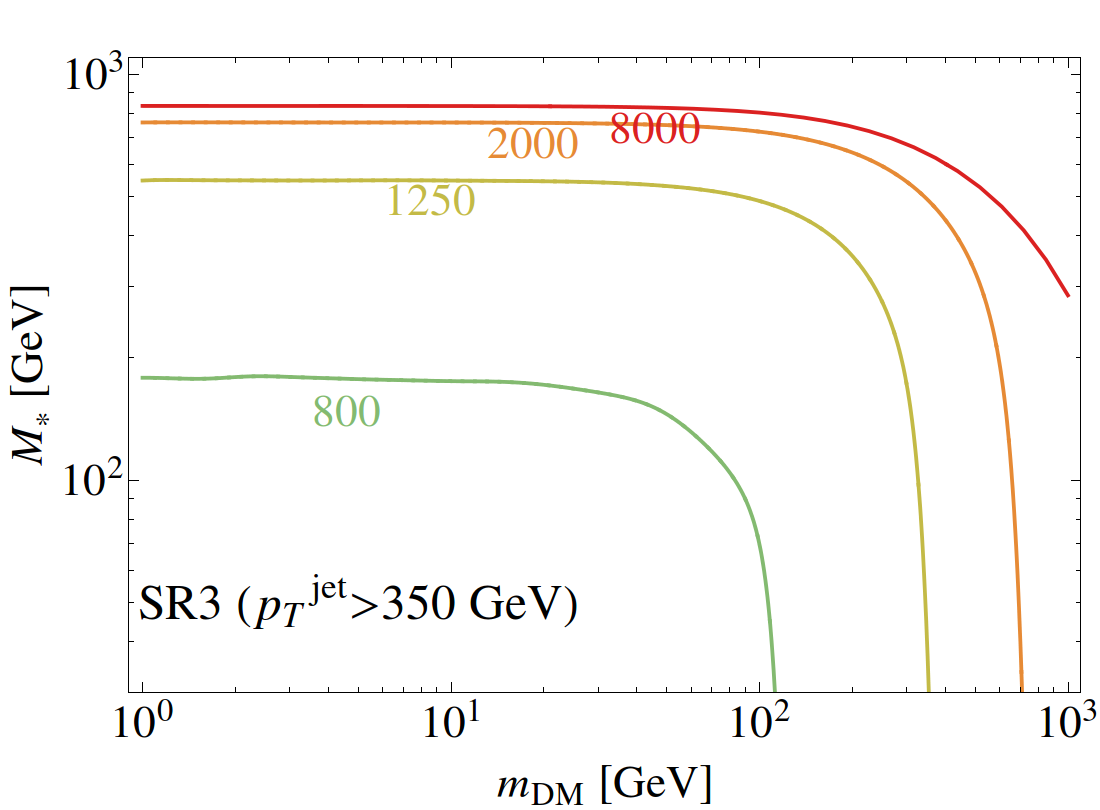}\hfill
  \includegraphics[width=0.49\textwidth]{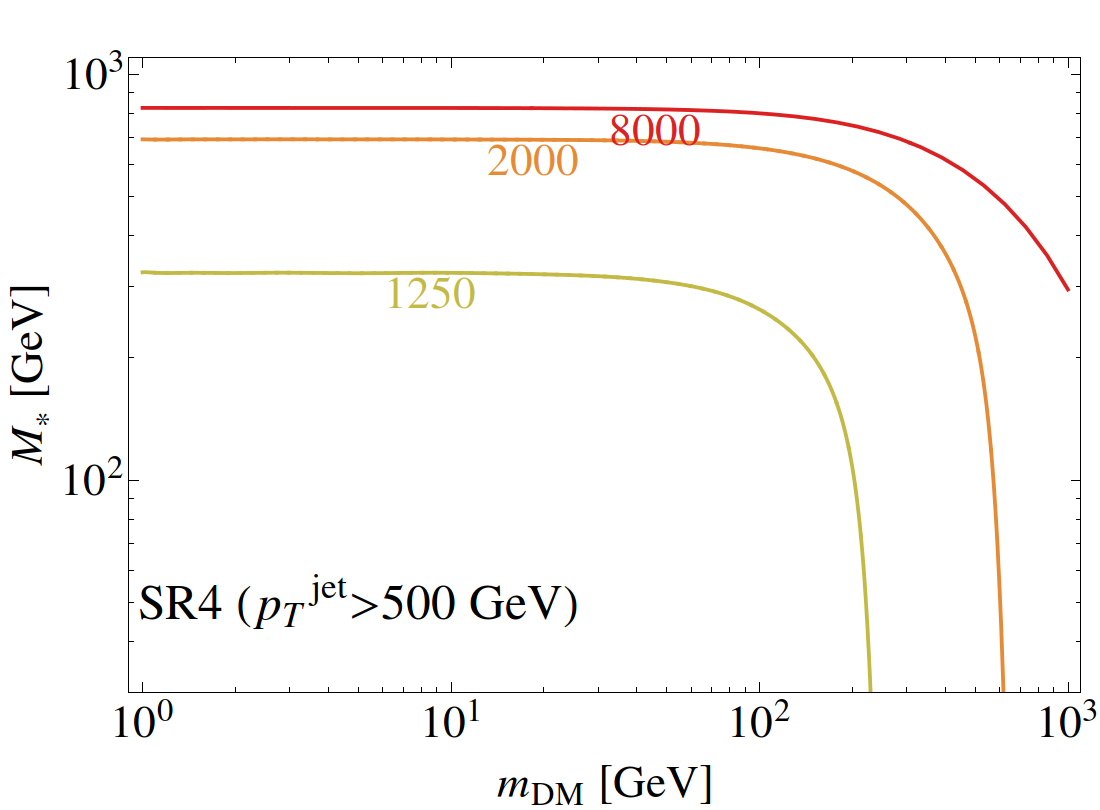}
  \caption{95\% c.l. lower bounds on $M_*$, as functions of $\MDM$, for some representative values of $\Mcut$ (in GeV), for the four signal regions of ref.~\cite{Aad:8tevmonojet}.}
 \label{fig:exclusion limits EFT fixed Mcut SR}
\end{figure}

At fixed $\MDM$ and $M_{\textrm{cut}}$, the ATLAS limits in eq.~(\ref{limits}) become lower bounds on the scale $M_*$, reported in fig.~\ref{fig:exclusion limits EFT fixed Mcut SR} as functions of $\MDM$ and for different values of $M_{\textrm{cut}} =$350, 450, 600, 800, 1250, 2000, 8000~GeV.
The four boxes in the figure correspond to the four different signal regions.  
The upper line in each plot,  $M_{\textrm{cut}} =$~8~TeV, corresponds to the na\"ive EFT limit, obtained without imposing any restriction on the centre-of-mass energy of the hard scattering.\footnote{The na\"ive EFT limit in SR$3$ differs from the ATLAS result on the \mbox{D$8$} operator by a $\sqrt[4]{2}$ factor, which reflects the factor $2$ enhancement of the cross-section for a Majorana DM particle with respect to the Dirac case considered in ref.~\cite{Aad:8tevmonojet}, if the same operator is used and the normalisation in eq.~(\ref{freeX}) taken into account.}
The limit deteriorates for decreasing $M_{\textrm{cut}}$ because of two distinct effects. 
The first one is that the total reference cross-section $\overline\sigma$ decreases, because it is restricted to a smaller kinematical range. 
This effect is unavoidable and ultimately due to the fact that the EFT cannot be trusted above its cutoff: trying to extrapolate the EFT above $M_{\textrm{cut}}$ would be inconsistent, and this is precisely why we restrict our signal to the $E_{\textrm{cm}} < M_{\textrm{cut}}$ region. 
The second effect is that the acceptances are also reduced, because the kinematical distributions of the restricted signal become softer, thus for decreasing $M_{\textrm{cut}}$ it becomes increasingly difficult to pass the cuts on $\pTjet$ and on $\MET$. 
Being dependent on the selection, this effect could be mitigated by softer cuts, compatibly with the minimal $\MET$ trigger requirement and with the fact that the SM background rapidly increases in the softer region. 
These considerations show that our signal is kinematically different from the na\"ive EFT prediction: an optimized limit in all $M_{\textrm{cut}}$ regions would require a dedicated study, which however goes beyond the scope of the present paper and can be properly performed only by the experimental collaborations. 

Going back to our results in fig.~\ref{fig:exclusion limits EFT fixed Mcut SR}, we notice that for large $M_{\textrm{cut}}$ the best limits are obtained from the SRs with harder cuts, namely from  SR$2$, SR$3$ and  SR$4$, which all have comparable reach. 
The low-cut region SR$1$ is instead not competitive with the other ones. 
The situation changes for low $M_{\textrm{cut}}$, because the cut acceptances decrease faster in the SR with harder cuts than in those with softer ones, and the limits start being dominated by the latter. 
For instance, when $M_{\textrm{cut}}$ goes below 500~GeV or so, the strongest $M_*$ bound starts coming from SR$1$, while the other SRs are no longer sensitive. 

The behaviour of the limits as functions of $\MDM$ is also easily understood. 
When $\MDM$ is lowered much below $M_{\textrm{cut}}$ and the kinematical cuts, the cross-section becomes independent of $\MDM$ and the limit saturates. 
The limit deteriorates as $\MDM$ increases, because the latter starts having a negative impact on the energy budget of the reaction.
The limit eventually disappears above a certain threshold, which corresponds to the region where the DM particle is too heavy to be produced with a centre-of-mass energy below $M_{\textrm{cut}}$. 
The minimal centre-of-mass energy is given by
\begin{equation}
\label{newtonwitten}
E_{\textrm{cm}}^{\textrm{min}} = \pTjet + \sqrt{\left(\pTjet\right)^2 + 4 \, \MDM^2} \,,
\end{equation}
where $\pTjet$ is the common jet and $\MET$ cut of each SR, out of which the mass threshold is then found to be~\footnote{The threshold effectively occurs for lower values of $\MDM$ when $M_{\textrm{cut}}$ gets close to the LHC threshold of $8$~TeV, because of the rapid large-x decrease of the parton distribution functions. }
\begin{equation}
\displaystyle
\label{newtonwittenweinberg}
\MDM^{\textrm{max}}= \frac{M_{\textrm{cut}}}{2} \, \sqrt{1 - 2 \, \frac{\pTjet}{M_{\textrm{cut}}} }  \, .
\end{equation}
We thus see once again that soft SRs are favoured for low $M_{\textrm{cut}}$, not only because they produce better $M_*$ limits, but also because they have an extended reach in the DM mass.\footnote{Formally, low $\pTjet$ improves the mass reach for any value of $\Mcut$. However, at large $\Mcut$ the threshold has a very poor sensitivity to the actual value of $\pTjet$ and all SRs have practically the same reach.}

\begin{figure}[t]
  \centering
  \includegraphics[width=0.49\textwidth]{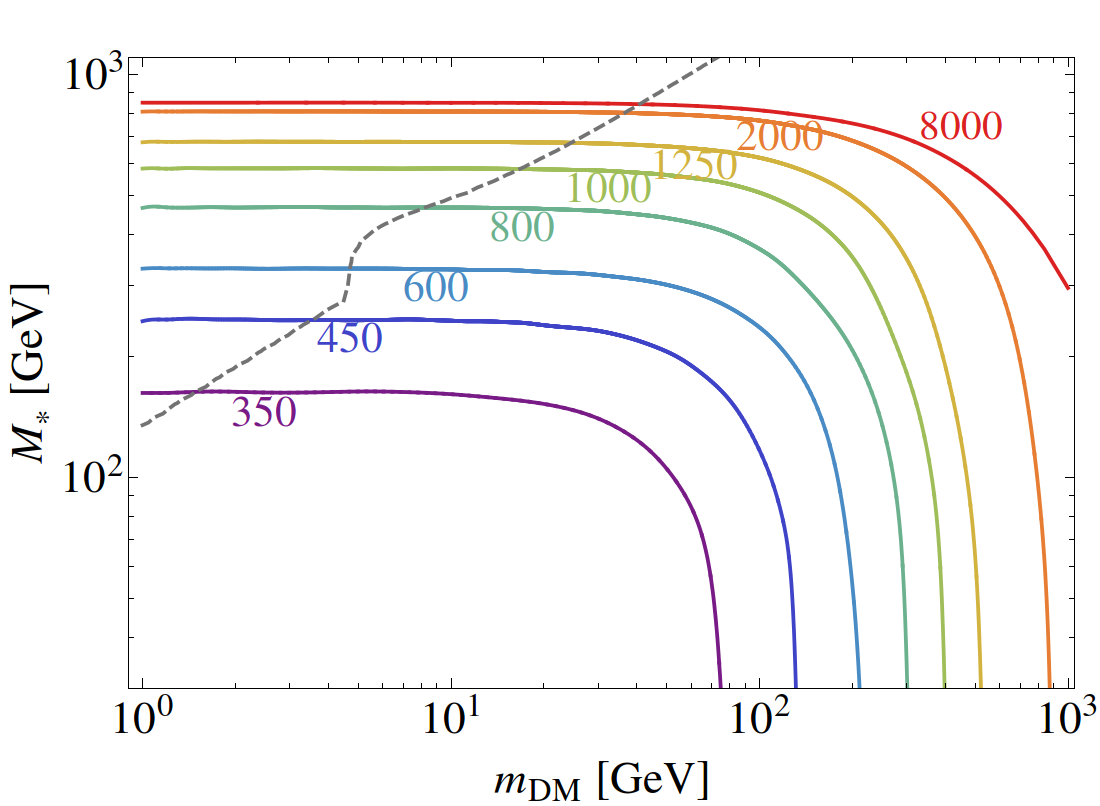} \hfill
    \includegraphics[width=0.49\textwidth]{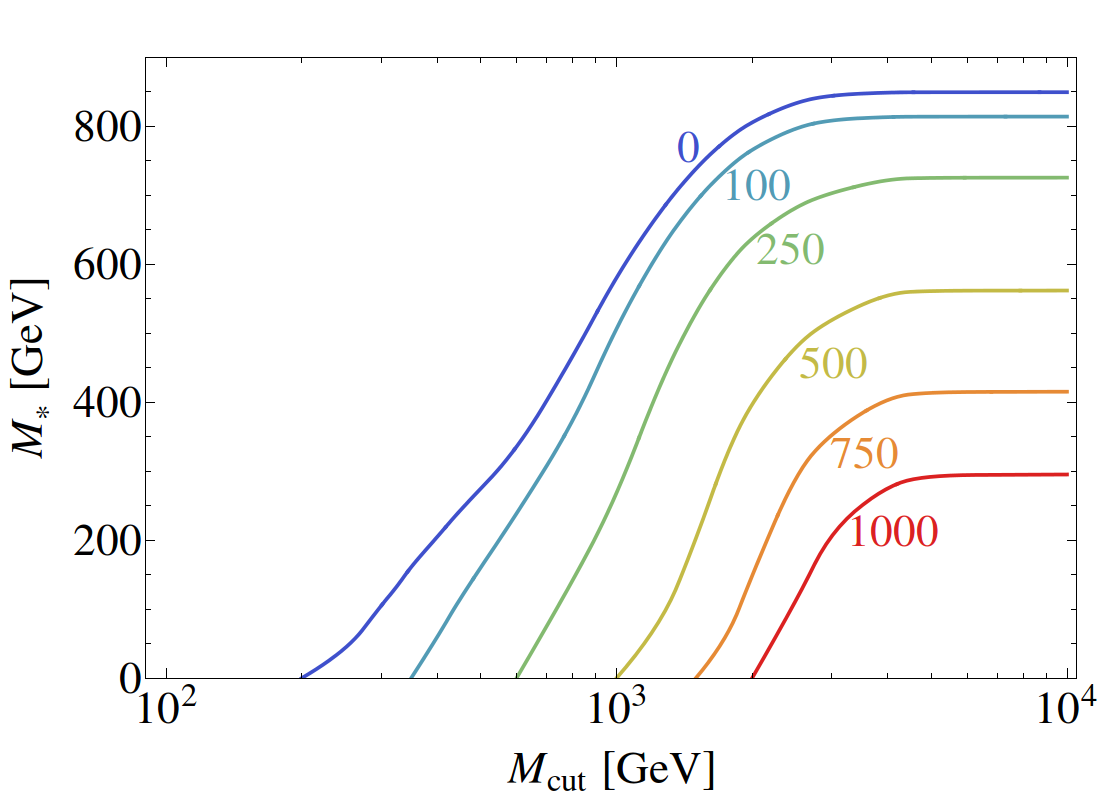} 
  \caption{Our combination of the lower bounds on $M_*$. {\em Left:} As a function of $\MDM$, for the same representative values of $\Mcut$ as in fig.~1. The dashed grey line is the relic density constraint. {\em Right:} As a function of $\Mcut$, for some representative values of $\MDM$ (in GeV).}
  \label{fig:joint limit MCut}
\end{figure}

The combined limits from all four SRs, obtained as the intersection of the allowed regions as described above or equivalently by taking the strongest $M_*$ bound at each point, are displayed in the left panel of fig.~\ref{fig:joint limit MCut}. 
The main conclusion we can draw is that the na\"ive EFT limit is fairly accurate when $\Mcut$ is significantly above $1$~TeV, while it considerably overestimates the actual exclusion for lower values of $\Mcut$. 
As an equivalent way to express the same information, the right panel of fig.~\ref{fig:joint limit MCut} shows the limit on $M_*$ as a function of $\Mcut$ for some fixed representative values of $\MDM$: 0, 100, 250, 500, 750, 1000~GeV. 
This representation is perhaps more convenient, as the dependence on $\MDM$ is rather smooth, and significant only in a limited range. 
Furthermore, it gives an idea of the search reach in the low $\Mcut$ region.
For reference, the dashed line on the left-hand panel of fig.~\ref{fig:joint limit MCut} shows the constraint from the relic density (under-abundant below the line and over-abundant above it), computed with the approximate analytical formulae for the EFT collected in Appendix~C.

The plots described above summarise the experimental situation in a simple and concise way, however they do not tell us how much of the theoretically allowed parameter space has been actually tested and how much is still unexplored. 
Namely, it is hard to establish a priori the ``reasonable'' $\Mcut$ values, and whether the corresponding $M_*$ limit should be regarded as a ``strong'' or a ``weak'' one.
We can do better if we remember that $\Mcut$ and $M_*$ are actually connected by eq.~(\ref{wittenpolchinski}).
Clearly, we do not know what $g_*$ is, but we do have some control on its value. 
We definitely know that it must be $g_* < 4\pi$, since taking it larger would make the EFT non-perturbative below the cutoff. 
This implies an upper bound on $\Mcut$ for any given $M_*$. 
In principle, there is no lower bound on $g_*$, it could be arbitrarily small pushing $\Mcut$ to smaller and smaller values. 
However, in a WIMP-like scenario we definitely expect $g_*\sim g_w \sim1$, to implement the WIMP miracle recalled in the introduction. 
Values of $g_*$ of order unity, and not radically smaller than that, should thus be considered as plausible benchmarks. 

\begin{figure}[t]
  \centering
  \includegraphics[width=0.65\textwidth]{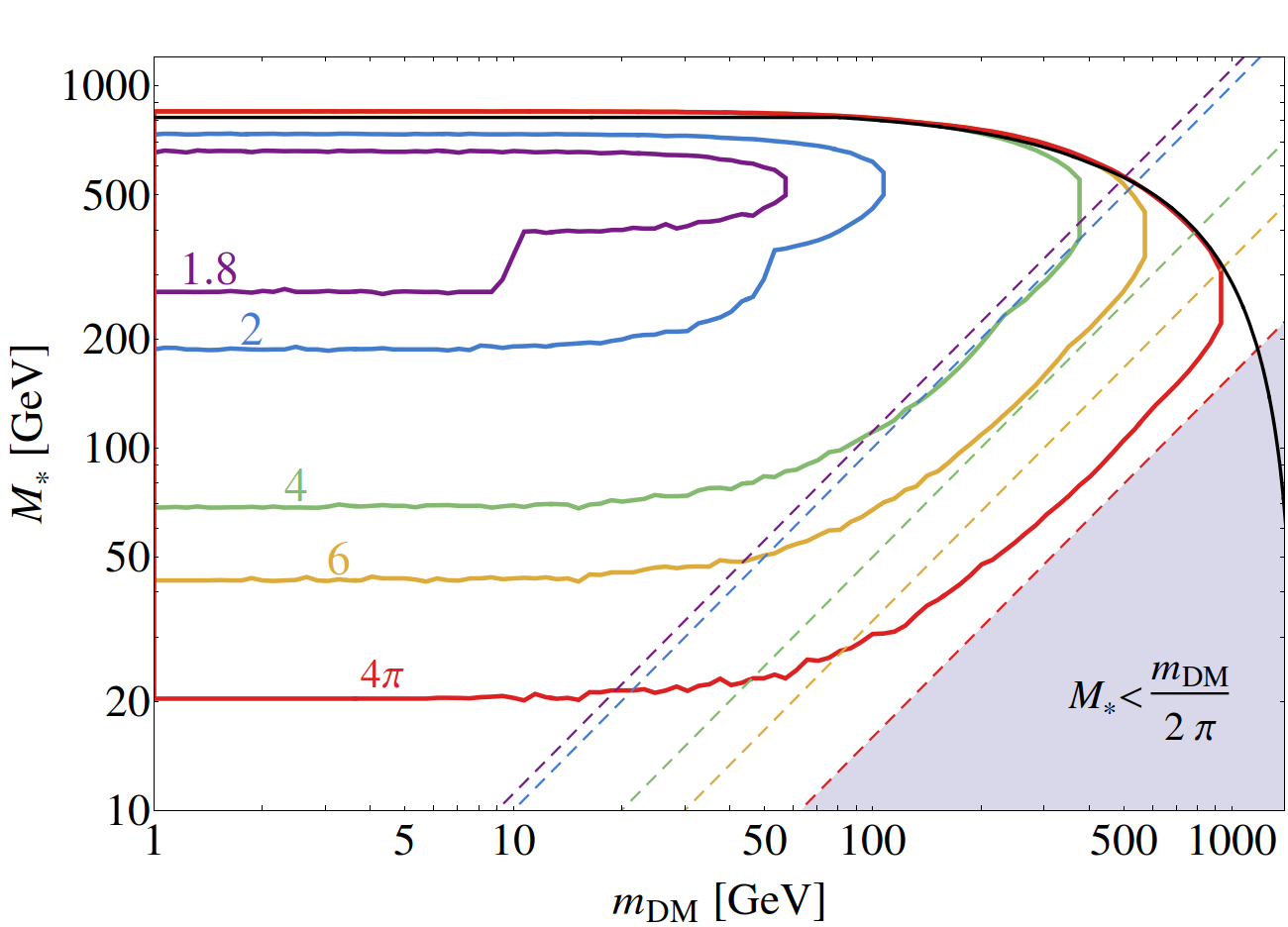}\vspace{10pt}
  \caption{The solid lines enclose the excluded regions in the plane $(\MDM,M_*)$, for some representative values of $g_*$, combining the four signal regions of ref.~\cite{Aad:8tevmonojet}. The black line is the limit one would obtain with the na\"ive EFT. The grey triangle is theoretically forbidden because of the self-consistency requirement $M_* > 2\MDM/g_*$, for $g_*=4 \pi$. The dashed lines show, with the same colour code as for the solid lines, how the grey triangle expands for smaller values of $g_*$.}
  \label{fig:exclusion limits EFT fixed gstar SR}
\end{figure}

The exclusion limits at fixed $g_*$, in the $(\MDM,\Mcut)$ plane, are shown by the coloured solid lines in fig.~\ref{fig:exclusion limits EFT fixed gstar SR}, for the representative values $g_*=1.8,2,4,6,4\pi$. The black solid line is the limit one would obtain in the na\"ive EFT.
We stress that closed excluded regions are obtained in this case, a fact that can be easily understood in the following terms. 
For a given $\MDM$, it is obvious that the limit must disappear at sufficiently large $M_*$, because the signal cross-section rapidly decreases for increasing $M_*$. 
However, the limit must also disappear for too low $M_*$, because at fixed $g_*$ lowering $M_*$ means lowering $\Mcut=g_*M_*$, which deteriorates and eventually kills the signal and the acceptances. 
There also exist values of $\MDM$ where these two competing effects do not allow to obtain an exclusion for any value of $M_*$, which is why the curves close on the right.
As a consequence, there are values of $g_*$ for which no limit on $M_*$ can be set, not even for $\MDM=0$. 
Our finding is quantitatively impressive: with the experimental results available so far, a satisfactory exploration of the parameter space has been possible only for $g_*$ above $4$ or $6$: the reference value $g_*=1$ is not excluded, and the smallest coupling we are sensitive to is $g_*\sim1.8$. 
Making progress in this direction would require more energy and integrated luminosity at the LHC, as expected in the forthcoming runs, but also improving the sensitivity to the small $\Mcut$ region as explained above. Indeed, the lower exclusion limits, in the low $\mDM$ region, are predicted by eq.~\eqref{newtonwitten} to occur near $g_* M_* = \Ecm^\text{min} \simeq 2 \pTjet$, where we take the lowest possible value for $\pTjet$, corresponding to 120 GeV for SR1 of \cite{Aad:8tevmonojet}.
This shows once again the importance of keeping the first signal region at the  lowest $\pTjet$ and $\MET$ values compatible with the trigger and background conditions.
As a last comment, we remind the reader that not all the points in fig.~\ref{fig:exclusion limits EFT fixed gstar SR} are theoretically allowed within the EFT framework. 
We are working here under the assumption of heavy-mediator DM, which means, as explained in the introduction, that $\MDM$ should be well below $\Mcut$, or at least $\MDM<\Mcut/2$, because otherwise there is no hope for the DM being produced within the range of validity of the EFT. 
This leads to the constraint $M_*=\Mcut/g_*> 2 \MDM/g_* $.  For $g_*=4 \pi$ this produces the grey theoretically forbidden region in fig.~\ref{fig:exclusion limits EFT fixed gstar SR}.
For $g_* < 4 \pi$ the boundary of the grey triangle moves as indicated by the dashed lines, with $g_*$ specified by the same colour code as for the solid lines.
However, eq.~\eqref{newtonwitten} guarantees that (in contrast with what we would obtain in the na\"ive EFT), the experimentally excluded region can at most approach the theoretically excluded one. 
Indeed, the closeness of the solid lines to the corresponding dashed lines gives a measure of how much the available EFT parameter space has been explored for the different values of $g_*$.
%


\section{Simplified model reinterpretation}
\label{simp_reint}

In the previous section we consistently derived from experimental data universal bounds on the EFT defined by the operator (\ref{eq:EFTlag}), as functions of the three relevant mass parameters $(M_*, \MDM, \Mcut)$. 
We now show how such bounds can be re-interpreted in any specific microscopic model underlying the chosen effective interaction. 
Since it collects only the contribution to the (positive-definite) signal cross-section coming from the kinematical region $E_{\textrm{cm}} < \Mcut$, where by definition the EFT is reliable, and it sets to zero the contribution corresponding to $E_{\textrm{cm}} > \Mcut$, our prescription for using consistently the EFT leads to underestimating the signal cross-section. We then expect our bounds to be systematically more conservative than those obtained by the direct comparison of a specific microscopic model with the experimental data. 
The aim of the present section is to perform a quantitative comparison of the limits derived with the two methods and to comment on the interpretation and practical consequences of any significant difference in the results. 

We consider two illustrative simplified models, characterized by quite different dynamics at the mediator scale, but nevertheless giving rise to the same leading effective operator (\ref{eq:EFTlag}) in the low-energy EFT.  
In Model~A, DM annihilation into quark-antiquark pairs and the inverse process occur via the $s$-channel exchange of a spin-1 $Z^{\, \prime}$ boson of mass $\mZp$, coupled to the axial-vector currents of quarks and DM with strengths $g_q$ and $g_X$, respectively. 
Very similar simplified models were discussed in refs.~\cite{Graesser:2011vj, An-s, APQ, LM, Hooper:2014fda, 1411.5917}.
In Model~B, the same processes occur via the $t/u$-channel exchange of color-triplet scalars of mass $\widetilde{m}$, with the same gauge quantum numbers as the squarks $\widetilde{q}$ of supersymmetric extensions of the SM, but with a universal Yukawa coupling of strength $g_{DM}$ to quarks and DM. 
Very similar simplified models were discussed in refs.~\cite{CEHL, An-t, DNRT, PVZ, GIRV, spanno}.
We have collected some useful details on the two models in Appendices~A and~B, respectively.   

\begin{figure}[t]
  \centering
  \includegraphics[width=0.75\textwidth]{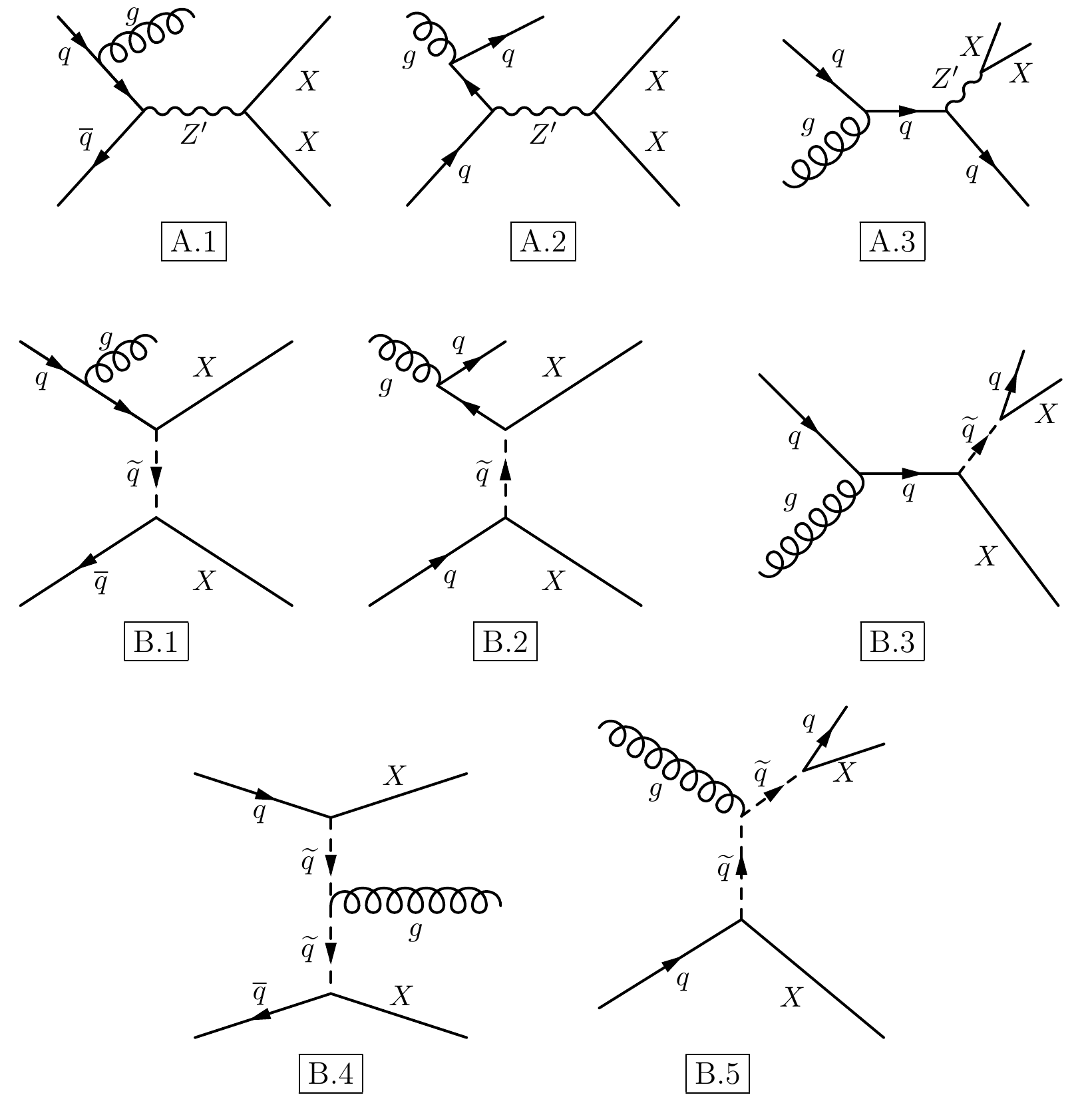}
\caption{Feynman diagrams describing the jet + $\MET$  DM signal at hadron colliders for models A ($Z'$ mediator) and B ($\widetilde{q}$ mediator) considered in the text.}
  \label{fig:diagrams}
\end{figure}

Before comparing the interpretation of the experimental results in the EFT and in the two simplified models, we display in fig.~\ref{fig:diagrams}
the tree-level Feynman diagrams contributing to the three hard partonic processes associated with the scattering $p p \rightarrow {\rm jet} + {\rm MET}$:
\begin{eqnarray}
 \text{(I)\,:  } & q(p_1) + \overline{q} (p_2) \rightarrow X(p_3) + X(p_4) + g(k) \, ;
\label{process1}
\\
 \text{(II)\,:  } & q(p_1) + g (p_2) \rightarrow X(p_3) + X(p_4) + q(k) \, ;
\label{process2}
\\
 \text{(III)\,:  } & \overline{q} (p_1) +  g(p_2) \rightarrow X(p_3) + X(p_4) + \overline{q} (k) \, .
\label{process3}
\end{eqnarray}
The symbols in brackets label the four-momenta of the corresponding particles.
Process~I is described by diagram A1 in Model A, by diagrams B1 and B4 in Model B. 
In the case of diagrams A1 and B1, it is understood that we should add the corresponding diagrams with the gluon radiated from the antiquark rather than from the quark line. 
Process~II is described by diagrams A2 and A3 in Model A, and by diagrams B2, B3 and B5 in Model B, plus those obtained by exchanging the momenta $p_3$ and $p_4$ of the Majorana DM fermion $X$.
Process~III is described by the same diagrams of process~II, with the prescription that all the arrows on the quark and squark lines should be reversed. 

The limits from our consistent EFT analysis and directly from the simplified models are obtained as follows. 
In the EFT, we compute the EFT parameters in each simplified model and we just apply the constraints derived in the previous section. 
The scale $M_*$ of the effective operator (\ref{eq:EFTlag})  is given by
\begin{equation}
\label{eq:MstarAB}
M_* = \frac{\mZp}{\sqrt{g_q \, g_X}} 
\quad
{\rm (Model~A)} \, ,
\qquad
\qquad
M_* = \frac{2 \, \widetilde{m}}{\gDM}
\quad
{\rm (Model~B)} \, .
\end{equation}
The cutoff scale $M_{\textrm{cut}}$, at which the EFT description loses its validity, is identified with the mediator mass $\Mmed$,  i.e. with $m_{Z'}$ in Model~A and with $\widetilde{m}$ in Model~B. 
Then, after this identification, the effective coupling $g_*$ is:
\begin{equation}
\label{eq:gstarAB}
g_* = \sqrt{g_q \, g_X} 
\quad
{\rm (Model~A)} \, ,
\qquad
\qquad
g_* = \frac{\gDM}{2}
\quad
{\rm (Model~B)} \, .
\end{equation}
To extract limits directly in the simplified models, we recast the ATLAS mono-jet analysis of ref.~\cite{Aad:8tevmonojet} as in section~\ref{ATMR}, with the only difference that now the signal cross-section is computed in the complete simplified model, i.e.\ with the diagrams in fig.~\ref{fig:diagrams} and with no $\Mcut$ restriction, for any value of $\Mmed$ and of $\MDM$.
For each point of the simplified model parameter space, the expected signal rate is computed in each SR and the corresponding exclusion limits are applied. 

\begin{figure}[t]
  \centering
  \includegraphics[width=0.49\textwidth]{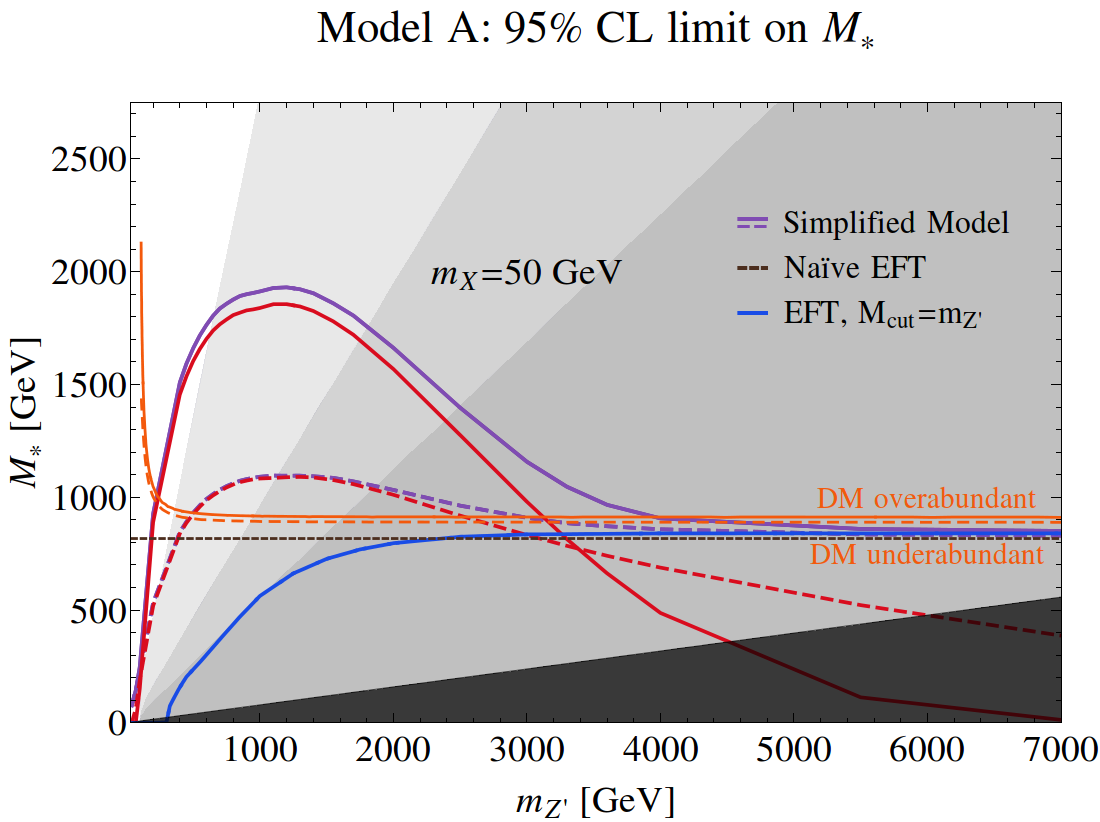} \includegraphics[width=0.49\textwidth]{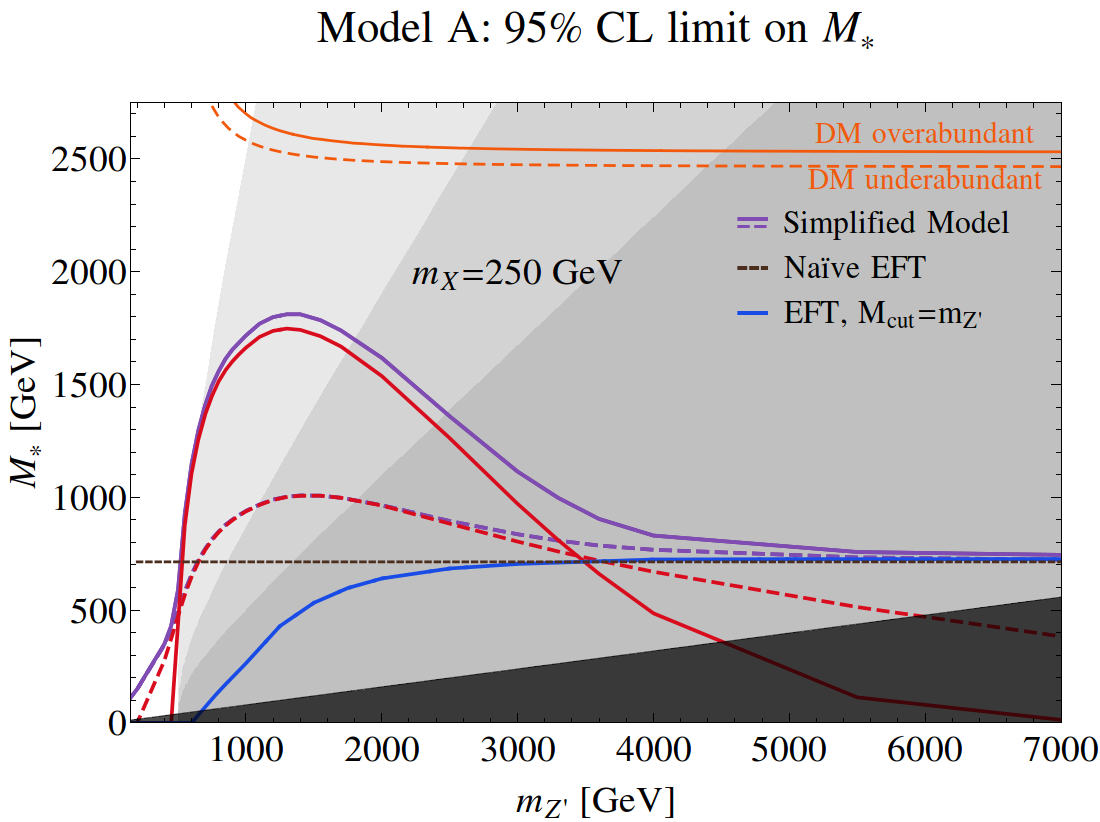}
 \caption{95\% CL limit on $M_*$ for Model A, as a function of $\mZp$, for $m_X = 50$ GeV (\textit{left})  and $m_X = 250$ GeV (\textit{right}). The horizontal dashed line corresponds to the limit obtained in the na\"ive EFT.  The blue line gives the limit consistently extracted in the EFT with $\Mcut=\mZp$.  All the other lines refer to the full model, and assume either $\GZp=\mZp/(8\pi)$ (solid) or $\GZp=\mZp/3$ (dashed). The purple lines show the limits obtained in the full model. The red lines corresponds to the resonant production of the mediator. The orange lines correspond to the correct relic abundance for a thermal freeze-out, computed according to the formulae for Model~A reported in appendix~C. From the top left to the bottom right, the increasingly dark grey shaded areas correspond to $\GZp/\mZp>1/(8 \pi),1/3,1$ and to $g_* > 4 \pi$.}
\label{fig:limits Model A}
\end{figure}

For Model~A, the result in the full model is illustrated by the purple lines in fig.~\ref{fig:limits Model A}, as an exclusion limit on $M_*$ as a function of $\Mmed\equiv\mZp$, for two representative values of $\MDM\equiv m_X$ and for two 
postulated values of the (width/mass) ratio of the mediator: $\Gamma_{Z'}/m_{Z'}=1/8\pi$ (solid) and $\Gamma_{Z'}/m_{Z'}=1/3$ (dashed). 
We will see below that using the $(\mZp,M_*)$ plane to represent the result suffers from an important limitation. 
Furthermore, $M_*$ is not a natural variable for the simplified model, where it is a derived quantity rather than a fundamental parameter. 
In this context, other ways of representing the limits could be more effective.
The choice of the $(m_{Z'},M_*)$ plane is however convenient for comparing these results with the EFT limits and with other studies of Model~A, such as those in refs.~\cite{1109.4398, BDM, Khachatryan:2014rra}.
In the figure, our consistent EFT limits, as reinterpreted in Model~A, are represented by blue solid lines, while the black dashed horizontal line shows the na\"ive EFT limit, formally obtained by sending $M_{\textrm{cut}}$ to infinity for fixed $M_*$.  For reference, the orange lines correspond to the correct relic abundance for  a thermal freeze out, computed here with the approximate analytical formulae for Model A reported in appendix C.
%

First, we can visually check that our consistent EFT limits are actually correct model-independent constraints, as they lie systematically below those obtained by working directly with the simplified model. 
Notice that this is not true for the na\"ive EFT limits, which overestimate the exclusion for very low mediator mass. 
Second, we observe that the limits obtained directly in Model~A are slightly stronger that the EFT ones, and that this effect is considerably amplified for a moderately light mediator in the case of the smaller $\Gamma_{Z'}/m_{Z'}$ ratio. 
The reason for this behaviour is that the simplified model cross-section can get significantly enhanced with respect to the EFT one, leading to a stronger bound, only thanks to the resonant production of the mediator, which can only take place if the latter is light enough. 
Furthermore, the resonant enhancement is of order $\pi\, m_{Z'}/\Gamma_{Z'}$, and this is why it is more pronounced for a narrow mediator. 
These considerations are made quantitative by the solid and dashed red lines in fig.~\ref{fig:limits Model A}, with the same conventions as before. 
These lines represent the limits on the simplified model obtained by computing the signal rate restricting the invariant mass of the $Z'$ propagators within two widths from its pole mass. 
The fact that the red lines are so close to the purple lines representing the ``true'' limit, when they are both significantly above the blue line, confirms that the resonant production is what drives the enhancement.
It also suggests that in this kinematical region DM searches in the simplified model should be actually regarded as mediator searches, and the results reported as limits on $\sigma (pp \rightarrow Z') \times$~BR($Z' \rightarrow XX$). Also, $Z'$ resonant production followed by the decay into quark-antiquark pairs, leading to a peak in the di-jet invariant mass distribution, may be a complementary signal to be looked for \cite{DY,Chatrchyan:2013qha, Aad:2014aqa, Khachatryan:2015sja}, with or without the extra jet: in such a case, we would obtain a limit on $\sigma (pp \rightarrow Z') \times$~BR($Z' \rightarrow q \overline{q}$). 
We will comment further on this in the conclusions.

We now turn to the aforementioned limitation of the $(m_{Z'},M_*)$ plane, which was already noticed in refs.~\cite{BDM, Khachatryan:2014rra, Aad:2015zva}, but we find important to emphasise. 
Model~A has four parameters: $\MDM$, $\mZp$, $g_q$, $g_X$. 
In fig.~\ref{fig:limits Model A}, the DM mass is set to a fixed value and each point of the plane uniquely determines $\mZp$ and $M_*$. Then also the product $g_q \, g_X$ is fixed by the left-hand side of eq.~(\ref{eq:MstarAB}), namely
\begin{equation}
\label{gqgx}
g_q \, g_X=\frac{m_{Z'}^2}{M_*^2}\,.
\end{equation}
Only one combination of the two couplings is left free at this point, and it might seem a good idea to fix it point-by-point to fit the values of $\Gamma_{Z'}/m_{Z'}$ that were assumed in drawing the purple lines in the figure. 
However, we must take into account that, for fixed $g_q g_X$, the accessible values of $\GZp/\mZp$ are bounded from below: 
\begin{equation}
\frac{\Gamma_{Z'}}{m_{Z'}} = \alpha \, g_q^2 + \beta \, g_X^2 \geq  g_qg_X \sqrt{4 \alpha \beta}=\frac{m_{Z'}^2}{M_*^2} \sqrt{4 \alpha \beta} \,,
\end{equation}
where $\alpha$ and $\beta$ are suitably defined coefficients (see appendix A) that do not depend on $g_q$ and $g_X$, and have only a mild dependence on the spectrum through phase space.
This means that the $(m_{Z'},M_*)$ plane is divided into regions, whose boundaries are curves (or, approximately, straight lines), where $\Gamma_{Z'}/m_{Z'}$ is always larger than a certain value. 
Some representative regions are displayed as grey shaded areas in fig.~\ref{fig:limits Model A}: from the top left to the bottom right, they correspond to $\Gamma_{Z'}/m_{Z'}>1/(8\pi),1/3,1$. 
The fourth and darkest region at the bottom right corresponds to $g_* = \sqrt{g_q g_X} >4 \pi$, where neither the EFT nor the simplified model admit a consistent perturbative description.   
In the neighbouring region where $\Gamma_{Z'}/m_{Z'}>1$, the EFT can still be consistently used, but the same does not apply to the chosen underlying simplified model: the fact that $\Gamma_{Z'}/m_{Z'}>1$ is telling us that in such strong coupling regime the simple mediator interpretation of the origin of the effective interaction breaks down. Even in the perturbative regime, the direct simplified model lines are obtained by assuming a given $\Gamma_{Z'}/m_{Z'}$, thus they become inconsistent on the right of the boundary of the corresponding $\Gamma_{Z'}/m_{Z'}$ region, because they cannot be associated to any physical point of the simplified model parameter space. 
On the left plot, for instance, we should have stopped drawing the purple and red solid lines corresponding to $\Gamma_{Z'} / m_{Z'} = 1/(8\pi)$ where they cross the boundary between the white and the very light grey region, at $\mZp \sim 600$~GeV. Similarly, we should have stopped the purple and red dashed lines, corresponding to  $\Gamma_{Z'} / m_{Z'} = 1/3$, where they cross the boundary of the two light grey regions, at $\mZp \sim 1.1$~TeV.
The only justification for keeping them is that the limits on the width are theoretical constraints, while the actual location of the curves is the result of the experimental analysis, which might improve its sensitivity in the future.
When this will happen the exclusion curves will move up and will exit more and more out of the inconsistent regions. 
As far as current data are concerned, however, this observation shows that the DM limits are actually rather poor, especially in the region of narrow mediator width, which corresponds to a weakly-interacting particle. 
But after all, this is exactly what we concluded from our exploration of the EFT parameter space: `small' $g_*$ effective couplings of order one are still unconstrained. 
Here we have just verified that the simplified model can not help us much in this respect.

\begin{figure}[h!]
  \centering
  \includegraphics[width=0.49\textwidth]{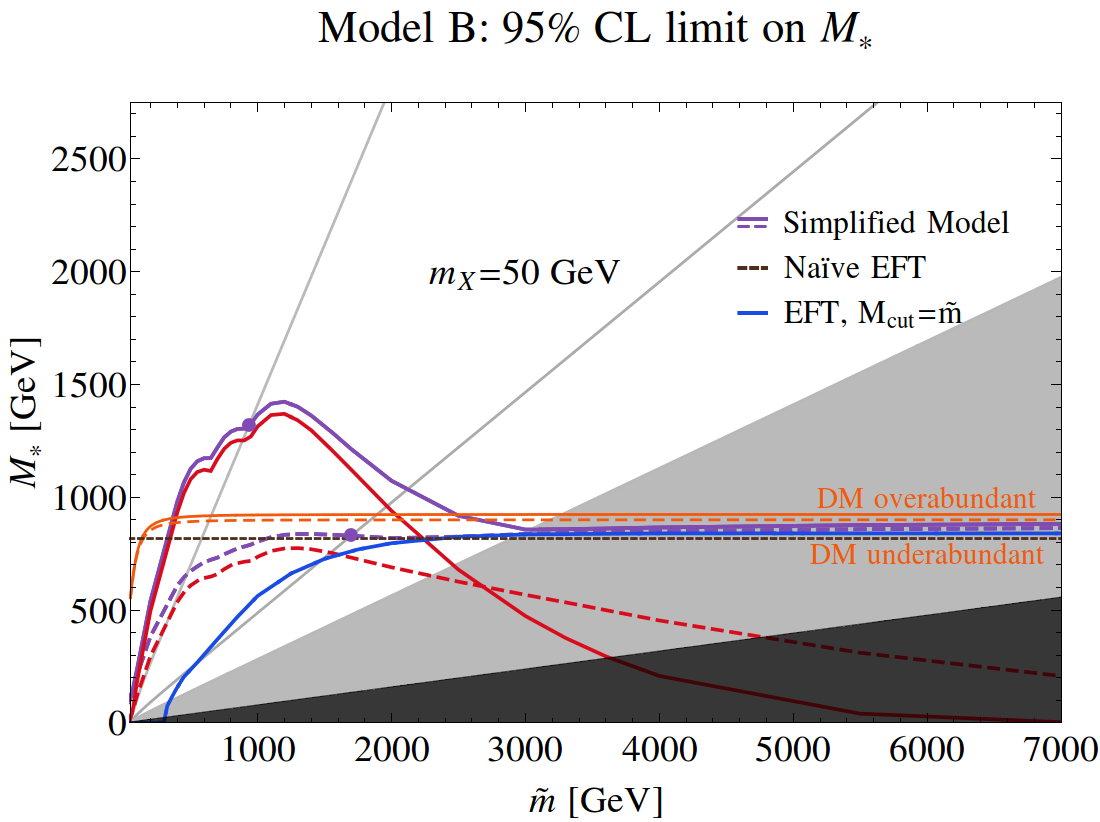}\vspace{10pt}
  \includegraphics[width=0.49\textwidth]{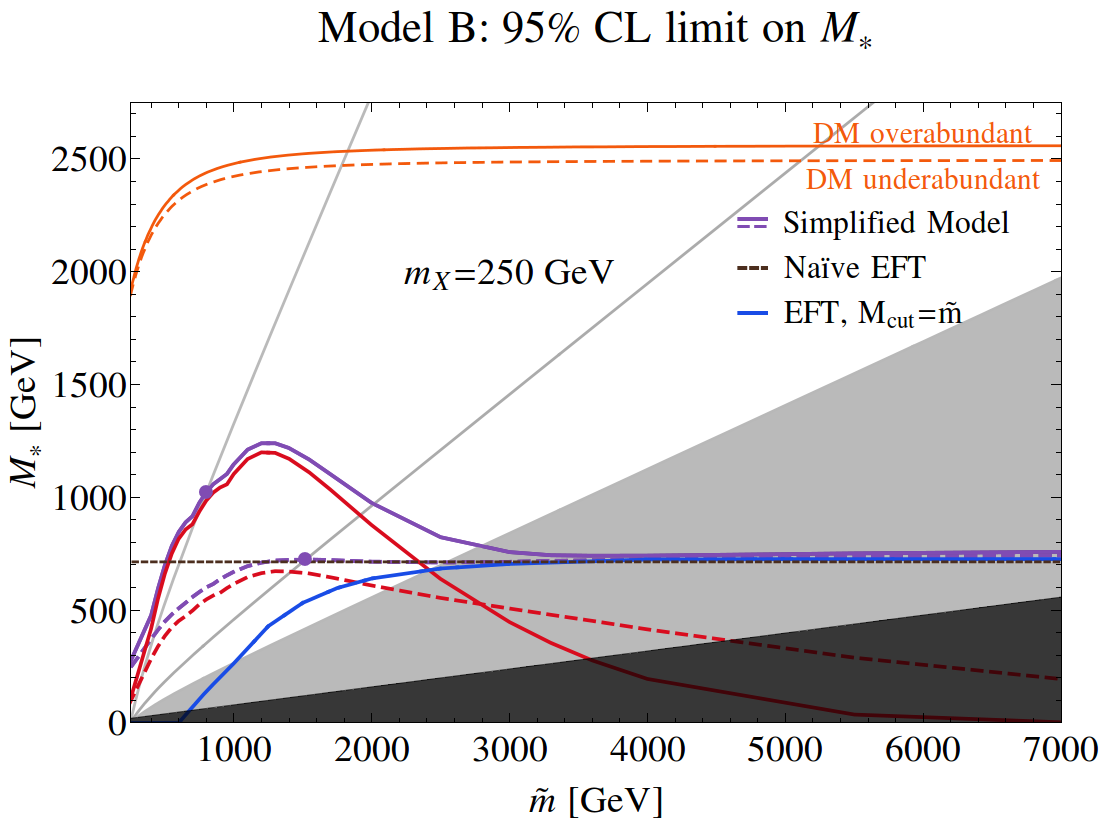}
  \caption{The same as in fig.~\ref{fig:limits Model A}, but for model B. The only difference is that, from top left to bottom right, the two diagonal lines correspond to $\Gamma_{\widetilde{q}}/{\widetilde{m}}=1/8\pi,\,1/3$, and the grey areas to $\Gamma_{\widetilde{q}}/{\widetilde{m}}>1$ and to $g_*=g_{DM}/2>4 \pi$.}
  \label{fig:limits Model B}
\end{figure}

Very similar considerations apply to Model~B, whose bounds are depicted in fig.~\ref{fig:limits Model B}. 
Also in this case the enhancement of the limit obtained directly in the simplified model is mostly due to the resonant production of the mediator, which can occur even in the so-called `$t$-channel mediator' case if an extra jet is emitted in the final state. 
This process corresponds (see diagrams B.3 and B.5 in fig.~\ref{fig:diagrams}) to an associated DM-$\widetilde{q}$ production followed by the $\widetilde{q}$ decay into DM plus jet. 
Once again, in the region of Model~B where the squarks are light enough, experiments should extend their selection criteria and look more generally for $n=1,2,\ldots$ jets plus $\MET$, to include the possibility of resonant squark production, both singly and in pairs. 
Of course, part of this is already being done in the context of standard squark searches within simplified supersymmetric models \cite{Chatrchyan:2014lfa, Aad:2014wea}. 
An experimental analysis along the above lines has been recently suggested by ref.~\cite{Maltoni} within a simplified supersymmetric model with a very light gravitino: something similar could be devised also for Model~B and similar simplified models for DM with `$t$-channel mediators'. 
A second point worth stressing for Model~B is that the issue with the $(\widetilde{m},M_*)$ plane is even more severe than in Model~A, because the model has only three parameters, therefore after fixing $\MDM$, $\widetilde{m}$ and $M_*$ the (width/mass) ratio of the mediator is fixed. 
In this case, fig.~\ref{fig:limits Model B} shows two lines corresponding to $\Gsq/{\msq}=1/8\pi,\,1/3$, a grey area where where $\Gsq>\msq$, and a dark grey area where $g_*= g_{DM}/2 > 4 \pi$. 
The only physical points of the four exclusion curves derived in Model~B (purple and red, solid and dashed) are those at the intersection with the lines corresponding to the assumed value of $\Gamma_{\widetilde{q}}/{\widetilde{m}}$, marked as full purple dots.

\subsection*{Other approaches} 
We are not the first to address the issues related with the na\"ive use of the EFT for DM in kinematical regimes extending beyond its range of validity: as already mentioned, they have been studied at length in the literature \cite{Fox, 1111.5331, riottos1, BDM, riottos2, riottot, Englert:2014cva, AABB, Malik}. 
In particular, refs.~\cite{riottos1, riottos2, riottot} proposed a criterion (recently adopted in refs.~\cite{Aad:2014tda, Aad:2015zva}) to estimate how sensitive the na\"ive limits on $M_*$ are to the unsafe region of the EFT and how much they deteriorate if the latter region is excluded from the analysis. 
Below we discuss two aspects of our approach in a way that can be helpful for the comparison with this previous literature.

The first point to be discussed concerns the choice of the kinematical variable to be used for discriminating the safe EFT region from the unsafe one. 
From the EFT viewpoint, the natural variable is clearly the hard scale of the process, $E_{\textrm{cm}}$: this was our choice for the present paper. 
However, within specific `mediator' models, or more precisely classes of models, another possible choice is the variable $\Qtr$, as proposed in refs.~\cite{riottos1, riottos2, riottot}.
$\Qtr=+\sqrt{|\Qtr^2|}$ is defined as the maximal virtuality of the mediator propagator, computed over the Feynman diagrams contributing to the partonic DM production process under study. 
Since $\Qtr < E_{\textrm{cm}}$, using $\Qtr$ to define the safe kinematical region of the EFT means gaining signal cross-section, thus obtaining a stronger and still reliable limit. 
Notice that, since the definition of $\Qtr$ depends on whether the mediator propagates in the $s$ or in the $t$ channel in the two-body annihilation $q \overline{q} \leftrightarrow XX$, $\Qtr$ is not suited for setting a model-independent limit.
However, one might still consider the two possibilities in turn and set separate limits for the two cases of $s$- and $t$-channel mediation. 
While this clearly does not exhaust all possibilities~\footnote{The effective interaction might well be generated by the combined exchange of $s$- and $t$-channel mediators, or by radiative effects not falling in any of these two categories.}, it might be still worth doing if it considerably enhances the reach.

To explore the exclusion reach of this method  and compare it with ours, we start by recalling the (trivial) expressions for $\Qtr$ in Models~A ($s$-channel) and B ($t$-channel), corresponding to the diagrams in fig.~\ref{fig:diagrams} and the conventions in eqs.~(\ref{process1})--(\ref{process3}). 
In Model~A, for both process I and process II (the kinematics of III is identical to that of II, so it does not need a separate discussion), $\Qtr$  is just the invariant mass of the DM pair
\begin{equation} 
\label{qtra}
\Qtr^2 = (p_3+p_4)^2 = (p_1+p_2-k)^2
\qquad \qquad
\text{(AI,\,AII)} \, .
\end{equation}
In model B, instead, we have to consider process I and II,III separately. In the case of process I, $\Qtr$ reads~\footnote{Notice that, if in Model B we had assumed a Dirac DM particle, only half of the conditions in eqs.~\eqref{qtrbI} and \eqref{qtrbII} should have been imposed. Therefore, the model dependence of this strategy depends on the assumptions made both on the mediator ($s$-channel or $t$-channel) and on the nature of the DM particle (Dirac or Majorana fermion, complex or real scalar).}
\begin{eqnarray} 
\Qtr^2 & = &  \max \left\{  
(p_1-k-p_4)^2=(p_3-p_2)^2 \, , \ 
(p_1-p_4)^2=(p_3-p_2+k)^2 \, , 
\right. \nonumber \\
& & 
\left.
(p_1-k-p_3)^2=(p_4-p_2)^2  \, , \
(p_1-p_3)^2=(p_4-p_2+k)^2
\right\} 
\qquad \text{(BI)} \, ,
\label{qtrbI}
\end{eqnarray}
while for process II,III we have
\begin{eqnarray} 
\Qtr^2 & = & \max \left\{  
(p_1-p_3)^2=(p_4-p_2+k)^2 \, , \
(p_3+k)^2=(p_1+p_2-p_4)^2 \, , \
\right. \nonumber \\
& & 
\left.
(p_1-p_4)^2=(p_3-p_2+k)^2  \, , \
(p_4+k)^2=(p_1+p_2-p_3)^2
\right\}  
\qquad \text{(BII)} \, .
\label{qtrbII}
\end{eqnarray}
Notice that the subprocesses are quantum-mechanically distinguishable and therefore it makes sense to adopt a different definition of $\Qtr$ for each of them.

\begin{figure}[thb]
  \centering
  \includegraphics[width=0.49\textwidth]{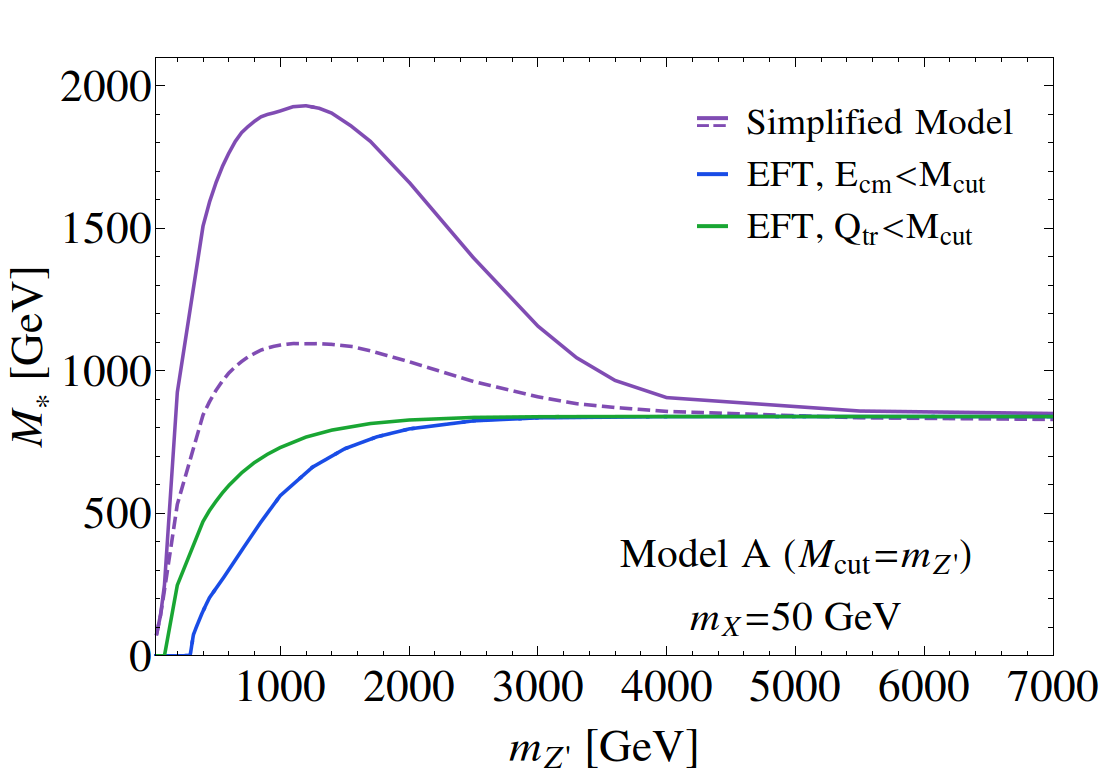}\hfill
  \includegraphics[width=0.49\textwidth]{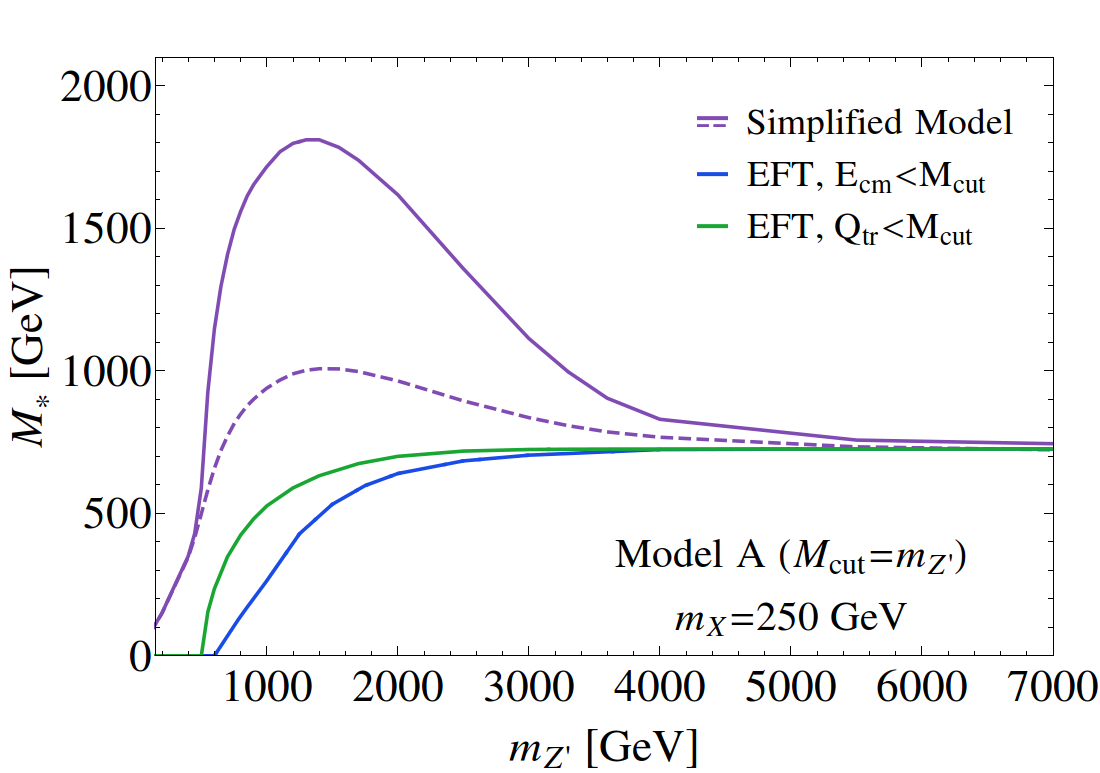}
  \includegraphics[width=0.49\textwidth]{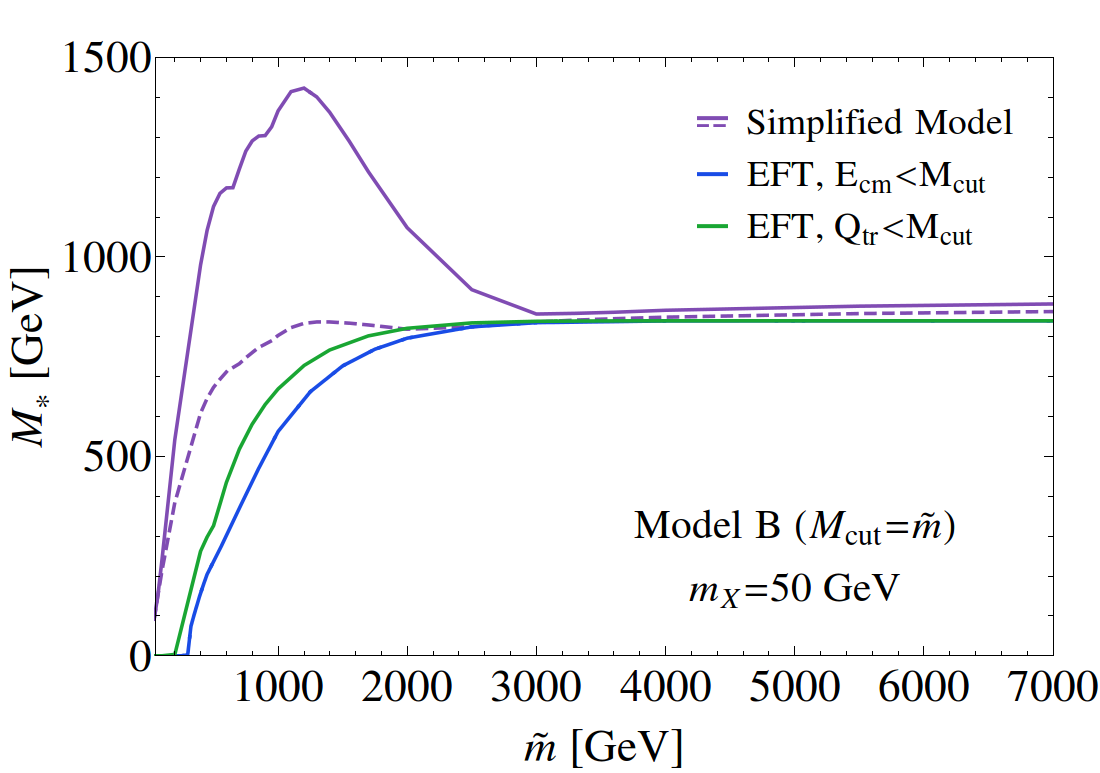}\hfill
  \includegraphics[width=0.49\textwidth]{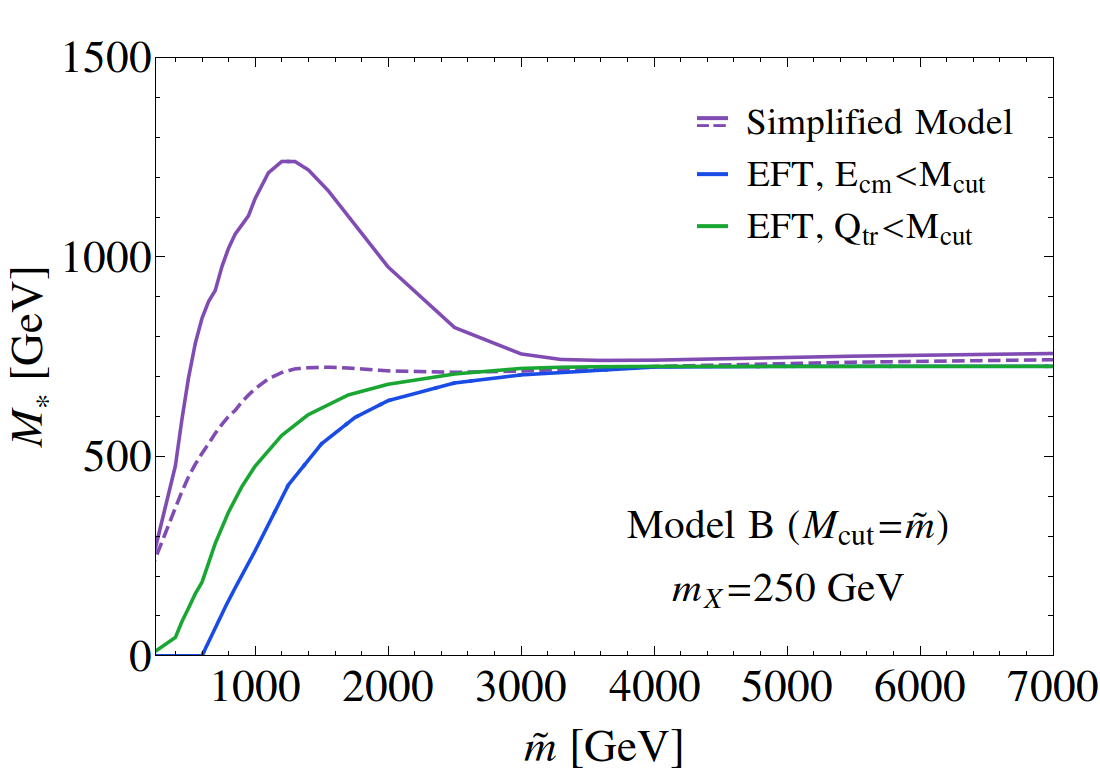}
 \caption{Limits on $M_*$ as functions of $\Mmed$ obtained for Models A and B with three different methods. 
The purple lines are derived in the full models, assuming two representative values of the ratio $\Gmed / \Mmed$: $1/(8\pi)$ (solid) and $1/3$ (dashed). 
The solid blue line is derived in the EFT with our method as described in the text. 
The solid green line is derived in the EFT by imposing the condition on $\Qtr$ proposed in refs.~\cite{riottos1, riottos2, riottot}.
{\em Upper plots:} Model A. 
{\em Lower plots:} Model B.}
\label{fig:limits comparison}
\end{figure}

The result of the comparison is displayed in fig.~\ref{fig:limits comparison}, where we show the limits on $M_*$ as functions of $\Mmed$, obtained for Models~A (upper plots) and B (lower plots) with three different methods. 
The purple and blue lines represent the full model and our approach to the EFT, respectively, namely the same curves as in figs.~\ref{fig:limits Model A} and \ref{fig:limits Model B}. 
The green line is also derived in the EFT, but with the cut $\Qtr<M_{\textrm{cut}}$ instead of $E_{\textrm{cm}} <\Mcut$. 
In the limit of heavy mediators, all the lines coincide as expected. 
The differences are in the region of relatively light mediators, where the EFT limit obtained with $\Qtr$  has, as expected, a better reach in $M_*$ than our method. 
However, in our view the improvement is not sufficiently significant, especially when compared with that obtainable in the full simplified model, to motivate the use of $\Qtr$ rather than $E_{\textrm{cm}}$. 
Our recommendation is thus to stick to the simple and model-independent version of our method, possibly trying to extend the reach by the direct search of the mediator which, as described in the previous section, is the sole responsible of the improved reach of the simplified model.

The second aspect to be clarified is that the consistent EFT limits in the $(\mDM,M_*)$  plane, at fixed $g_*$, cannot be inferred from those obtained in the na\"ive EFT by just performing a rescaling of $M_*$. 
It is indeed clear that such a rescaling cannot lead to closed exclusion curves such as those we obtained in fig.~\ref{fig:exclusion limits EFT fixed gstar SR}. 
One might be tempted to consider a rescaling here because the EFT cross-section is proportional to $1/M_*^4$, so that the reduction of the cross-section caused by the kinematical cut might be reabsorbed into an effective $M_*$. 
Namely, one might consider defining the ratio~\footnote{Using $\Qtr$ or $E_{\textrm{cm}}$ makes no difference for the point we want to make here.}
\begin{equation}
\label{eq:R ATLAS}
R (M_*,\mDM,g_*) =\frac{\sigma(M_*)\Big\lvert_{\Qtr<\Mcut=g_*M_*}}{\sigma(M_*)}\, ,
\end{equation}
where $\sigma$ denotes the signal cross-section computed in the na\"ive EFT for a given signal region. 
At fixed $g_*$ and $\mDM$, $R$ is a function of $M_*$, which tends to one for sufficiently high $M_*$ and to zero for sufficiently low $M_*$, because of the effect of the kinematical cut illustrated in eq.~(\ref{newtonwitten}). 
Given that $R$  measures the reduction of the cross-section with respect to the na\"ive EFT, one might think of getting the limit on $M_*$ at each $\mDM$, call it $\Mresc$, starting from the limit obtained in the na\"ive EFT, call it $\Mblind$, and solving the implicit equation
\begin{equation}
\label{eq:rescaling of Mblind}
\Mresc = \left[ R\left(\Mresc,\mDM\right)\right]^\frac 14 \Mblind\,.
\end{equation}
The effective operator scale $\Mresc$ obtained in this way is the one that reproduces, in the EFT with the cut on $\Qtr$, the same signal cross-section that was needed for setting the bound at $\Mblind$ in the na\"ive EFT.
Namely, eq.~(\ref{eq:rescaling of Mblind}) is equivalent to
\begin{equation}
\label{eq:rescaling of Mblind simpler}
\sigma(\Mresc)\Big\lvert_{\Qtr<\Mcut=g_*\Mresc} = \sigma(\Mblind)\,,
\end{equation}
where we have exploited the fact that in the na\"ive EFT $\sigma(M_*)$ simply scales as $1/M_*^4$. 

\begin{figure}[ht]
  \centering
  \includegraphics[width=0.65\textwidth]{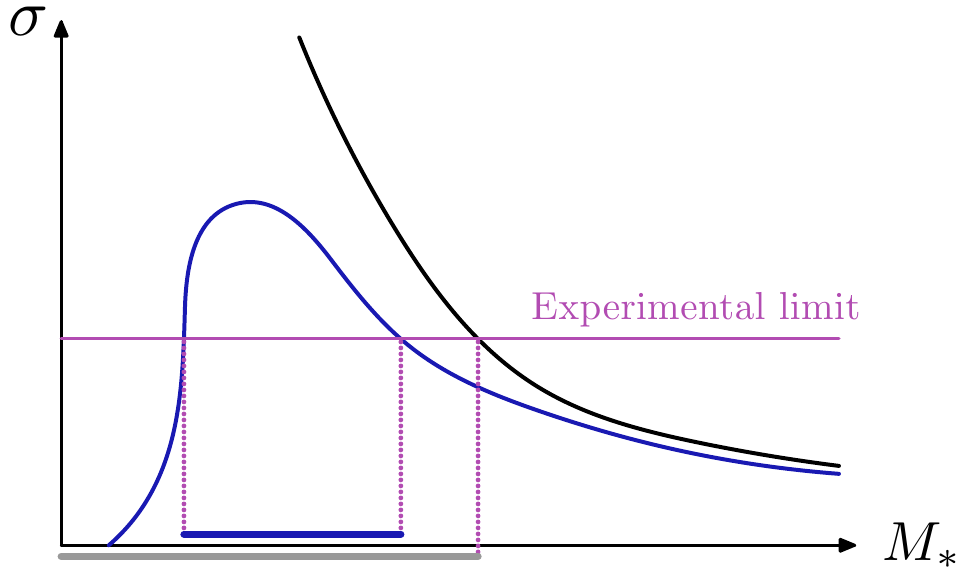}
  \caption{A pictorial representation of how the bounds on $M_*$ depend on the prescription for computing the signal in the EFT. The signal cross-section is displayed as a function of $M_*$, for fixed $g_*$ and $\MDM$. 
The black and the blue lines correspond to the na\"ive EFT and to our consistent prescription, respectively. 
The horizontal purple line represents the experimental limit. The resulting excluded interval for $M_*$ is reported near the horizontal axis for the two prescriptions.}
\label{fig:pedagogical}
\end{figure}

The above method for obtaining $\Mresc$ is more elaborate than directly comparing the experimental limit on the cross-section with the prediction of the kinematically restricted EFT, as we suggested in section~2.1.
Furthermore, the rescaling method might obscure the fact that eq.~(\ref{eq:rescaling of Mblind simpler}), or equivalently eq.~(\ref{eq:rescaling of Mblind}), has either zero (which means that no limit can be set) or two solutions for $\Mresc$, but it never has only one. 
The behaviour of the restricted EFT cross-section, compared with the na\"ive EFT, is pictorially represented in  fig.~\ref{fig:pedagogical}. 
The cross-section vanishes before approaching $M_*=0$, because of the cut $\Qtr<g_* M_*$. 
Therefore there are two values of $M_*$ for which the cross-section equals the experimental limit, which means that the excluded region has one upper but also one lower limit in $M_*$, differently from the na\"ive EFT as depicted in the figure. 
Therefore, the true limit cannot be set by just rescaling the na\"ive EFT exclusion curve. 
Notice also that certain strategies to solve eq.~(\ref{eq:rescaling of Mblind}), such as applying an iterative procedure starting from $\Mresc=\Mblind$, might obscure the existence of the lower bound, as they systematically converge to the upper one. 
The quantitative impact on the excluded regions in the $(\MDM,M_*)$ plane, for different values of $g_*$, was already displayed in fig.~\ref{fig:exclusion limits EFT fixed gstar SR} for our kinematical requirement $\Ecm<g_*M_*$.

\section{Conclusions}

We described a simple strategy to set robust and model-independent limits on heavy-mediator DM at the LHC. 
Our method is based on the generic form of the operators in the EFT containing only the DM and the SM particles, with no assumptions on the underlying dynamics. 
However, it also takes into account the presence of a cutoff scale $M_{\textrm{cut}}$ above which the EFT loses its validity. 
$\Mcut$ must be regarded as one of the free parameters of the EFT, on the same footing as the DM mass $\MDM$ and the effective interaction scale $M_*$. 
We have to do so if we aim at a comprehensive exploration of the whole range of theoretical possibilities. 
The parameter $M_{\textrm{cut}}$ can be traded for $g_*$,  the typical coupling strength at the mediator scale.
As explained in the paper, $g_*$ can be defined in the EFT alone, and further characterised for any assumed underlying model.

We applied our method explicitly to the ATLAS mono-jet search of ref.~\cite{Aad:8tevmonojet}, obtaining the exclusion contours in the $(\MDM,M_*)$ plane shown in fig.~\ref{fig:exclusion limits EFT fixed gstar SR}, for fixed representative values of $g_*$. 
We believe that this kind of plots illustrates the current experimental situation in an accurate and comprehensive way, providing a fair assessment of the LHC sensitivity to heavy-mediator DM. 
At the moment, we are only sensitive to large values of $g_*$, while the region $g_*\sim1$, which is arguably the most natural one for WIMP DM, is still largely untested. 
Making progress requires higher energy and luminosity, but also an optimisation of the experimental search strategies. 
As pointed out in section~2, our signal is kinematically different from that of the na\"ive EFT, in particular it is characterised by softer $\MET$ and $\pTjet$ distributions. 
The reach of the searches would then benefit from a sensitivity improvement in the soft region.

In section~\ref{simp_reint} we considered two different simplified models, both giving rise to the same effective operator considered in section~2. 
We compared our EFT limits, reinterpreted in the two models, with those obtained from a dedicated comparison of the experimental bounds with the prediction of the two models, reaching two main conclusions. 
First, the limits set within the simplified models can be considerably stronger than the EFT ones, but only because of the resonant production of the mediator, which enhances the simplified model cross-section. 
Therefore, a DM search performed within a simplified model (in the only interesting region where the limit is potentially stronger than in the EFT) is actually not a search for DM, but a search for the mediator, and as such it should be interpreted. 
The canonical \mbox{$\sigma\times$BR} limit as a function of the mediator mass appears to be the best option for presenting the experimental results.
The second conclusion is that the current experimental sensitivity is still rather poor, even when working within a simplified model. 
In particular the region of weak coupling, i.e. narrow mediator, is mostly unexplored, in accordance with what we found in our EFT analysis.
We finally discussed two aspects of our approach, to facilitate the comparison with the recent literature. We found that the usage of the variable $\Qtr$ in place of $E_\text{cm}$ to define the safe kinematical region of the EFT  does not improve the sensitivity significantly enough to pay back for the increased model-dependence.  We also remarked that just rescaling the na\"ive EFT limit does not account for the impossibility, within mono-jet searches, of excluding arbitrarily low values of $M_*$ at fixed $\MDM$ and $g_*$.

In summary, we have found that the LHC sensitivity to the heavy-mediator DM hypothesis is still limited and wide regions of the parameter space still wait to be explored. On the experimental side, improving the analysis in the soft region would be of great help. On the phenomenological side, more comprehensive methods should be elaborated to cover each different region of the parameter space with the most suitable strategy. Non-resonant DM signals are well described by the EFT which, as outlined in the present paper, when consistently used provides a robust model-independent way to approach the problem. Within specific models, this needs to be supplemented by resonant mediator searches, which however should be performed by exploiting fully the predictive power of the assumed mediator dynamics. This means taking into account all the mediator production mechanisms (single and/or pair) and all its possible decay modes, including the one to visible objects which might give complementary informations. These aspects are left to future work.

%
%
\acknowledgments
We thank A.~De~Simone, M.~de~Vries, E.~Morgante, T.~Jacques, G.~Polesello, S.~Rahatlou, A.W.~Riotto, F.~Riva, A.~Romanino, A.~Urbano, and M.~Zanetti for discussions. We  acknowledge {\em Heidi} for computational support. This work was supported in part  by the ERC Advanced Grant no.267985 (\textit{DaMeSyFla}), by the MIUR-PRIN project 2010YJ2NYW and by the MIUR-FIRB Grant RBFR12H1MW.
%
%
%
%
%
\appendix
%
\section{Model A: axial-vector mediator}
\label{MA}
We collect here some details on the first of the two simplified models considered in the text, Model A.
Previous discussions of very similar models can be found in refs.~\cite{Graesser:2011vj, An-s, APQ, LM, Hooper:2014fda, 1411.5917}.
The mediator is a neutral vector boson $Z^{\, \prime}$, singlet under the SM gauge group, with mass $m_{Z^{\, \prime}}$, a universal axial coupling $g_q$ to quarks, no renormalisable couplings to leptons, and an axial coupling $g_X$ to the Majorana DM fermion $X$ of mass $m_X$.
Since the model is introduced for purely illustrative purposes, without making reference to an underlying more fundamental theory, we introduce an explicit $Z'$ mass term and we neglect $Z$-$Z^{\, \prime}$ mixing, as well as anomalies and their cancellation mechanisms.

The model Lagrangian is
\begin{eqnarray}
\label{LA}
{\cal L}_A & =  & {\cal L}_{SM} + {\cal L}_{X} + {\cal L}_{Z^{\, \prime}} + {\cal L}^A_{int} \, , \\
{\cal L}_{Z^{\, \prime}} & =  & -\frac 14 Z^{\, \prime}_{\mu\nu} Z^{\, \prime \, \mu\nu} + \frac 12 \mZp^2 Z^{\, \prime}_\mu Z^{\, \prime \, \mu} \, , \\
{\cal L}^A_{int} & = & Z^{\, \prime}_\mu \left( g_q \sum_q \overline q \gamma^\mu\gamma^5 q + g_X \overline X \gamma^\mu\gamma^5 X \right) \equiv Z^{\, \prime}_\mu \, J_{Z^{\, \prime}}^\mu  \, ,
\label{LAint}
\end{eqnarray}
where ${\cal L}_{SM}$ is the SM Lagrangian,  ${\cal L}_X$ is the free Lagrangian for $X$ in (\ref{freeX}), $Z^{\, \prime}_{\mu\nu} = \partial _\mu Z^{\, \prime}_\nu-\partial _\nu Z^{\, \prime}_\mu$, and the sum in (\ref{LAint}) runs over all quark flavours ($q=u,d,c,s,t,b$). The model has four parameters, 
\begin{equation}
\label{Apar}
g_q \, , 
\quad
g_X \, ,
\quad
\mZp \equiv \Mmed \, ,
\quad
m_X \equiv \MDM \, , 
\end{equation}
which can all be taken to be real and positive (in principle, $g_q$ and $g_X$ could have either sign, but this is not relevant for the present study).

Notice that the choice of a purely axial interaction, universal for all quark flavours, is crucial to generate the effective interaction (\ref{eq:EFTlag}) from (\ref{LAint}) in the low-energy limit. At leading order in $E/\mZp \ll 1$, the approximate solution of the $Z^{\, \prime}$ equations of motion is
\begin{equation*}
Z^{\, \prime \, \mu} = -\frac{1}{\mZp^2} J_{Z^{\, \prime}}^\mu \, ,
\end{equation*}
which substituted in (\ref{LA}) gives
\begin{eqnarray}
{\cal L}_{EFT}^A = & - & \frac{g_X^2}{2\mZp^2} (\overline X \gamma^\mu \gamma^5 X)(\overline X \gamma_\mu \gamma^5 X)  \label{eq:A EFT XXXX}  \\
& - & \frac{g_q^2}{2\mZp^2} \sum_q (\overline q \gamma^\mu \gamma^5 q)\sum_q (\overline q\gamma_\mu \gamma^5 q) \label{eq:A EFT qqqq}  \\
 & - & \frac{g_qg_X}{\mZp^2} (\overline X \gamma^\mu \gamma^5 X)\sum_q (\overline q\gamma_\mu \gamma^5 q)\, . \label{eq:A EFT XXqq}
\end{eqnarray}
The effective interaction term  (\ref{eq:A EFT XXqq}) between the SM quarks and the DM field reproduces the one in (\ref{eq:EFTlag}) as long as 
\begin{equation}
\label{eq:Mstar model A}
M_* = \frac{\mZp}{\sqrt{g_q \, g_X}} \, .
\end{equation}
Notice also that integrating out the heavy $Z^{\, \prime}$ generates two additional four-fermion operators, (\ref{eq:A EFT XXXX}) and  (\ref{eq:A EFT qqqq}). However, \eqref{eq:A EFT XXXX} is subject only to very mild constraints from the limits on DM self-interactions. The four-quark operator \eqref{eq:A EFT qqqq} can be probed by the searches for contact interactions \cite{Dreiner:2013vla, 1409.4657}, but can be parametrically suppressed by choosing $g_X>g_q$ for fixed $g_*$.

At tree-level, and including only two-body decays, the total decay width of the $Z'$ is
\begin{equation}
\label{eq:width_mediator_A}
\GZp = 
\frac{\mZp}{12\pi}\left[ 2g_X^2\left(1-\frac{4m_X^2}{\mZp^2} \right)^{3/2} +\sum_q 3g_q^2\left(1-\frac{4m_q^2}{\mZp^2} \right)^{3/2} \right] \, ,
\end{equation}
with the obvious modifications if some of the final states are not kinematically accessible. 

%
%
%
\section{Model B: coloured scalar mediators}
\label{MB}
We collect here some details on the second of the two simplified models considered in the text, Model B.
Previous discussions of very similar models can be found in refs.~\cite{CEHL, An-t, DNRT, PVZ, GIRV, spanno}.
In Model B, the interactions between the SM quarks and the DM particle X are mediated by three families of degenerate complex scalars of mass $\widetilde{m}$, with the same gauge quantum numbers of the corresponding left- and right-handed quarks. 
Since they are identical to the squarks of supersymmetric extensions of the SM, we denote them with the same symbols,  ($\widetilde{u}_{iL}, \widetilde{d}_{iL}, \widetilde{u}_{iR}, \widetilde{d}_{iR}$), where $i=1,2,3$ are family indices. 
Similarly, the Majorana fermion $X$ mimicks, although in the simplified fashion specified by its interactions below, the lightest neutralino of supersymmetric models.

The model Lagrangian reads
\begin{eqnarray}
\label{LB}
{\cal L}_B & =  & {\cal L}_{SM} + {\cal L}_{X} + {\cal L}_{\widetilde{q}} + {\cal L}^B_{int} \, , \\
{\cal L}_{\widetilde{q}} & =  & 
\sum_{i=1}^3 \left[
(\partial^\mu \widetilde{u}_{iL})^\dagger (\partial_\mu \widetilde{u}_{iL}) 
+
(\partial^\mu \widetilde{d}_{iL})^\dagger (\partial_\mu \widetilde{d}_{iL}) 
+
(\partial^\mu \widetilde{u}_{iR})^\dagger (\partial_\mu \widetilde{u}_{iR}) 
+
(\partial^\mu \widetilde{d}_{iR})^\dagger (\partial_\mu \widetilde{d}_{iR}) 
\right.
\nonumber \\
\label{Lsq}
& - &
\left. 
\widetilde{m}^2  \left( 
\widetilde{u}_{iL}^{\, \dagger} \widetilde{u}_{iL} +
\widetilde{d}_{iL}^{\; \dagger} \widetilde{d}_{iL} +
\widetilde{u}_{iR}^{\, \dagger} \widetilde{u}_{iR}+
\widetilde{d}_{iR}^{\; \dagger}  \widetilde{d}_{iR} 
\right) \right] + \ldots \, , \\
{\cal L}^B_{int} & = & 
- g_{DM} \left[ \sum_{1=1}^3
 \left(
\widetilde{u}_{iL} \, \overline{u_{iL}} 
+ 
\widetilde{d}_{iL} \, \overline{d_{iL}} 
+ 
\widetilde{u}_{iR} \, \overline{u_{iR}} 
+ 
\widetilde{d}_{iR} \, \overline{d_{iR}}  
 \right) 
X + {\rm h.c.} \right]
\, ,
\label{LBint}
\end{eqnarray}
where ${\cal L}_{SM}$ and ${\cal L}_X$ are the same as in Model A, and the dots in (\ref{Lsq}) denote the squark gauge interactions, generated by promoting ordinary derivatives to SM covariant derivatives. 
Notice that the mass degeneracy and the universality of the Yukawa couplings among quarks, squarks and DM evade the typical problems of supersymmetric models with flavour-changing neutral currents. 
The model has three parameters, 
\begin{equation}
\label{Bpar}
g_{DM} \, ,
\quad
\widetilde{m} \equiv \Mmed \, ,
\quad
m_X \equiv \MDM \, , 
\end{equation}
which can all be taken to be real and positive ($g_{DM}$ can be complex, but it can be chosen to be real and positive by absorbing its phase into a redefinition of the squark fields).

As for Model A, we can derive the EFT by solving the classical equations of motion for the squarks in the low-energy limit $E \ll \widetilde{m}$: 
\begin{equation}
\widetilde{u}_{iL} = - \frac{\gDM}{\widetilde{m}^2} \, \overline X u_{iL} \, ,
\quad
\widetilde{u}_{iR} = - \frac{\gDM}{\widetilde{m}^2} \, \overline X u_{iR} \, ,
\quad
\widetilde{d}_{iL} = - \frac{\gDM}{\widetilde{m}^2} \, \overline X d_{iL} \, ,
\quad
\widetilde{d}_{iR} = - \frac{\gDM}{\widetilde{m}^2} \, \overline X d_{iR} \, .
\end{equation}
Substituting into ${\cal L}_B$ yields
\begin{eqnarray}
{\cal L}_{EFT}^B & = &  \frac{\gDM^2}{\widetilde{m}^2} \sum_{i=1}^3 
\left[ 
(  \overline X  u_{iL} ) (\overline{u_{iL}}  X) 
+
(  \overline X  u_{iR} ) (\overline{u_{iR}}  X) 
+
(  \overline X  d_{iL} ) (\overline{d_{iL}}  X) 
+
(  \overline X  d_{iR} ) (\overline{d_{iR}}  X) 
\right]
\nonumber
\\
& = &
- \, \frac{\gDM^2}{4 \, \widetilde{m}^2} 
\,
\left(  \overline X \gamma^\mu \gamma^5 X \right)
\, 
\left[ \sum_{i=1}^3 
\left( \overline{u_i} \gamma_\mu \gamma^5 u_i 
+ \overline{d_i} \gamma_\mu \gamma^5 d_i  \right) \right] \, , 
\label{eq:B EFT lagrangian}
\end{eqnarray}
where for the second equality we have used the Fierz identities and the fact that when $X$ is a Majorana spinor $\overline{X} \gamma^\mu X =0$.
The effective interaction term (\ref{eq:B EFT lagrangian}) between the SM quarks and the DM particle reproduces the one in (\ref{eq:EFTlag}) as long as 
\begin{equation}
M_* = \frac{2 \, \widetilde{m}}{g_{DM}} \, .
\label{eq:Mstar model B}
\end{equation}

At tree-level, and assuming $\widetilde{m} > m_X + m_q$, where $q$ is the corresponding quark, the decay width of the generic $\widetilde{q}$ is
\begin{equation}
\label{eq:width_mediator_B}
\Gsq = \frac{\msq}{16\pi}\gDM^2 \sqrt{1+\frac{(m_q^2+m_X^2)^2}{\msq^4}-2\frac{m_q^2+m_X^2}{\msq^2}}\left(1-\frac{m_q^2}{\msq^2}-\frac{m_X^2}{\msq^2}\right) \,.
\end{equation}
%

%
%
\section{Formulae for the relic density}
We collect here the approximate analytical formulae used for the calculation of the relic density in the EFT (fig.~2), in Model A (fig.~5) and in Model B (fig.~6), before requiring that it reproduces the recent precise determination by the Planck collaboration \cite{Planck} (for our purposes, the latter can be rounded to $\ODM h^2 = 0.12$ with negligible error). They can be straightforwardly derived from the existing literature (see e.g. \cite{kolb1990, gondolo1991}). Up to terms of order $1/x_f$, where $x_f$ is the value of $x=m_X/T$ at freeze-out:
\begin{equation}
\label{eq:relic abundance}
\ODM h^2 \approx 1.07\cdot 10^{9}\, (\textrm{GeV})^{-1} \ \frac{x_f}{\sqrt{g_*}\, \mP \frac{1}{16\,m_X^2}\left(a + \frac{3b}{x_f}    
\right)}\,,
\end{equation}
where $h$ is the dimensionless Hubble parameter, $g_{*} \sim 100$ is the number of relativistic degrees of freedom, $\mP \simeq 2.4 \times 10^{18} \, {\rm GeV}$ is the reduced Planck mass, $m_X$ is the mass of the DM particle in GeV, and 
\begin{eqnarray}
x_f & = & \ln(\lambda)-\frac 12 \ln[\ln(\lambda)] + \ln\left[1+ 
\frac{6b}{a} 
\frac{1}{\ln(\lambda)} \right] \, ,\\
 \lambda & = & 0.038 \frac{2}{\sqrt{g_*}}\mP\, m_X \left(\frac{a}{16\, m_X^2}\right)\,.
\end{eqnarray}
In the EFT, introducing the dimensionless parameters $\alpha_q = m_q / m_X$, 
\begin{eqnarray}
a & = & \sum_q \frac{96}{\pi} \left( \frac{m_X}{M_*} \right)^4 \alpha_q^2  \, \sqrt{1-\alpha_q^2} \,, \\
b & = & \sum_q \frac{4}{\pi}  \left( \frac{m_X}{M_*} \right)^4 \left(8 -16 \alpha_q^2 +11 \alpha_q^4 \right)
 \left(1-\alpha_q^2 \right)^{-1/2}\, ,
\label{eq:EFT a b} 
\end{eqnarray}
where the sums run over the quark flavours whose mass is below $m_X$. 
 
In the two models underlying the EFT, we introduce two additional dimensionless parameters, $\beta=m_X/\Mmed$ and $\gamma=\Gamma_{\rm med}/\Mmed$, to account for the finite mass $\Mmed$ and width $\Gamma_{\rm med}$ of the mediator. Then in Model A ($Z'$ mediator)
\begin{eqnarray}
a &=&  \sum_q \frac{96}{\pi} g_q^2g_X^2 \frac{\beta^4 \sqrt{1-\alpha_q^2}}{(4\beta^2-1)^2+\gamma^2}\ \alpha_q^2\left(1-8\beta^2+16\beta^4\right)\,, \\
b & = & \sum_q \frac{4}{\pi} g_q^2g_X^2 \frac{\beta^4}{\sqrt{1-\alpha_q^2}\left[(4\beta^2-1)^2+\gamma^2\right]^2} \left\{ (8-16\alpha_q^2+11\alpha_q^4)(1+\gamma^2) \right. \nonumber \\
&& -  8\beta^2\left[ (8-16\alpha_q^2+14\alpha_q^4)+3 \alpha_q^2 (2 -\alpha_q^2) \gamma^2  \right] \nonumber \\
&& +  16\beta^4\left[ (8-16\alpha_q^2+26\alpha_q^4)+3 \alpha_q^2 (4 -3\alpha_q^2) \gamma^2 \right] \nonumber \\
&& +  \left. 768 \,\beta^6 (\beta^2-1) \, \alpha_q^4 \right\}\, ,
\label{eq:A a b} 
\end{eqnarray}
and in Model B ($\widetilde{q}$ mediator)
\begin{eqnarray}
a & =&\sum_q \frac{6\gDM^4}{\pi}\frac{\beta^4 \sqrt{1-\alpha_q^2}}{(1+\beta^2-\alpha_q^2\beta^2)^2+\gamma^2}\, \alpha_q^2\,,\\
b & = & \sum_q\frac{\gDM^4}{4\pi}\frac{\beta^4}{\sqrt{1-\alpha_q^2} \left[ (1+\beta^2-\alpha_q^2\beta^2)^2+\gamma^2\right]^3}
\left\{
(8-16\alpha_q^2+11\alpha_q^4)(1+\gamma^2)^2 \right. \nonumber \\ &
& +  4 \beta^2(1-\alpha_q^2)(4-18\alpha_q^2+11\alpha_q^4)(1+\gamma^2) \nonumber \\&
& +  2\beta^4 (1-\alpha_q^2)^2 [(8-48\alpha_q^2 +33\alpha_q^4)+(8-24\alpha_q^2+11\alpha_q^4) \gamma^2] \nonumber \\&
& +  4\beta^6(1-\alpha_q^2)^3 (4-10\alpha_q^2+11\alpha_q^4) \nonumber \\&
& +  \left. \beta^8(1-\alpha_q^2)^4(8+11\alpha_q^4) \right\} \, .
\label{eq:B a b} 
\end{eqnarray}
%
%
%
%

\end{document}